# Mechanical Studies of an Additional Light Baffle for the LSST Camera

Hannah Mary Margaret Pollek*[a], Gabriele Rodeghiero[b], John Andrew[c], Alex Drlica-Wagner[d], Alessio Taranto[b,e], Luca Rosignoli[e], Aashay Pai[f], Douglas R. Neill[c], Travis Lange[a], Andrew P. Rasmussen[g], Aaron Roodman[g], Pierre Antilogus[h], Alexandre Boucaud[i], Martin Nordby[a]

[a]SLAC National Accelerator Laboratory, Menlo Park, CA, United States
[b]INAF - Osservatorio di Astrofisica e Scienza dello Spazio, Bologna, Italy
[c]Vera C. Rubin Observatory, Tucson, AZ, United States
[d]Fermi National Accelerator Laboratory, Batavia, IL, United States
[e]Department of Physics and Astronomy (DIFA), University of Bologna, Bologna, Italy
[f]Department of Physics, University of Chicago, Chicago, IL 60637, United States
[g]Kavli Institute for Particle Astrophysics and Cosmology, SLAC National Accelerator Laboratory, Menlo Park, CA, United States
[h]Institut National de Physique Nucléaire et de Physique des Particules (IN2P3) du CNRS, Paris, France
[i]Université Paris Cité, CNRS, Astroparticule et Cosmologie, Paris, France

*Author contact information: pollek@slac.stanford.edu

## ABSTRACT

Commissioning the NSF-DOE Vera C. Rubin Observatory consisted of engineering operation and on-sky data-taking, initially with the Commissioning Camera followed by the commissioning run of the LSST Camera (LSSTCam). As with other wide-field astronomical projects, the Rubin team anticipated a significant amount of stray light effects which would necessitate investigation and systematic mitigation. This led the Rubin stray light working group to develop tools, including a robust model of the entire observatory in Zemax, to trace the light paths of stray light artifacts back to their sources. This model along with the other efforts of the working group enabled significant improvements in stray light mitigation leading up to the commencement of the Legacy Survey of Space and Time (LSST).

One such potential source was identified as a small chamfer on the L3 lens, for which it was hypothesized that a simple baffle added inside of the LSSTCam near the L3 should prove beneficial to the quality of data being collected in the LSST. Initial Zemax models proved this hypothesis to be correct, but it is important to weigh the improvements made versus the effort, risk, and cost especially when considering any hardware modifications to an instrument that is already running and collecting immense amounts of data each night.

This paper investigates the impacts of installing an L3 baffle via a collection of mechanically focused studies, where the principal areas of focus are installation feasibility, baffle geometry, materials & coating selection, and potential impacts to the purge system.

**Keywords:** LSST, LSSTCam, Rubin, Stray Light, Baffle, Optics, Telescope, Ray Tracing

## 1. INTRODUCTION

The NSF-DOE Vera C. Rubin Observatory, funded by the U.S. National Science Foundation and the U.S. Department of Energy's Office of Science, has officially begun collecting the largest dataset of astronomical data in history, the Legacy Survey of Space and Time. Prior to the commencement of operations, it first underwent a rigorous testing and engineering data collection period to refine the observatory's ability to image the cosmos. During this dress-rehearsal of the survey, observing scientists practiced their roles, settings were fine-tuned, and as with most observatories, stray light artifacts were observed [1].

As these features were spotted, the team began immediately working on identification and mitigation of each one, for the benefit of the immense amount of data to be collected in this decadal survey. Some of the features were traced back to originating from reflections or refractions off of hardware within the LSSTCam. The LSSTCam is the world's largest digital camera, with a 3.2 gigapixel resolution, three lenses, and a set of filters used interchangeably to image the cosmos.

The L3 lens, which is the closest of the three lenses to the focal plane, sitting just behind the online filter has a chamfered internal edge that has been identified as a source of stray light artifacts.

Stray light artifacts have a direct impact on the results of astronomical surveys by muddling the images and potentially altering the photometry of astronomical features. Thus, it is vital to the Rubin team to mitigate stray light as completely as possible while maintaining the etendue of the instrument. This paper will highlight the process of identifying and characterizing two specific stray light artifacts that originate from the L3 chamfer and then will delve in detail into the mechanical feasibility and impact studies surrounding the proposed addition of a new baffle into the LSSTCam.

## 2. STRAY LIGHT ARTIFACT CHARACTERIZATION

To understand how we might mitigate any given stray light feature, we must first investigate the feature and determine where it is propagating from, and in what conditions it appears. Each stray light artifact that is observed by the Rubin team gets identified, grouped with others believed to be from the same origin, and then named for ease of identification.

The first artifact that is pertinent to this paper is referred to as the "Angel Wings" due to the morphology of the feature resembling feathered wings coming in from the edges of the focal plane as seen in Figures 1 & 2. They were later re-named to be the "L3 Angel Wings" after their source was discovered to be a feature on the third lens (L3) in the LSSTCam.

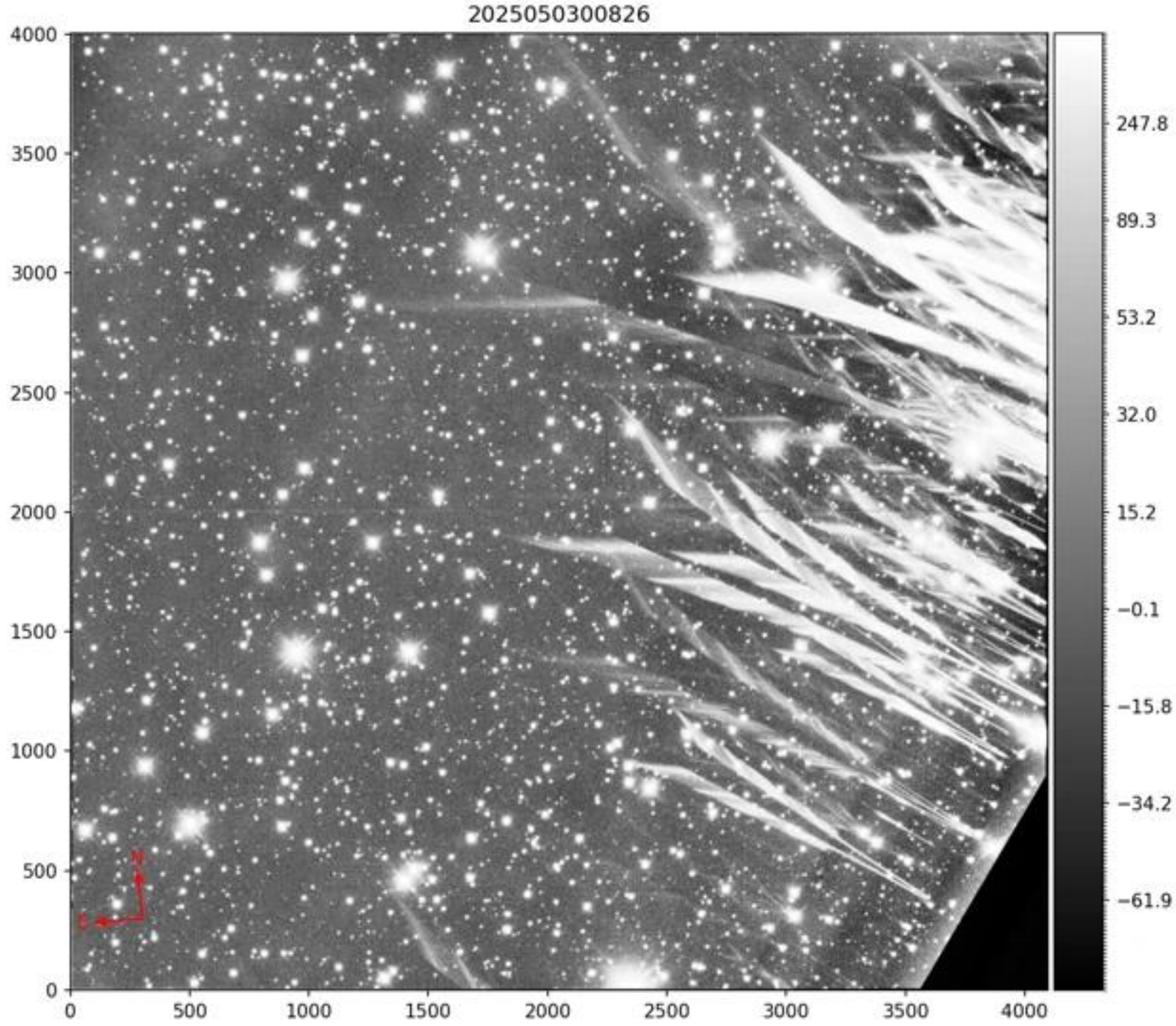

Figure 1: Example of "L3 Angel Wings" artifact at the right side of the image here, taken in May of 2025 during the commissioning period of the observatory (visit = 2025050300826). This image is from a single CCD, which is about 0.23 x 0.23 degrees on sky.

The L3 Angel Wings were first spotted in an image taken by the LSSTCam in April of 2025, during early on-sky observations in the engineering/testing period. They appear in the corners or edges of the field of view in the off-axis occurrence range of 1.99 degrees to 2.1 degrees, just slightly outside the 1.75 degree FOV. Also in April of 2025, the Zemax simulation developed by the Rubin stray light working group reproduced these features, at which point ray tracing was performed (see Figure 3), determining the origin of L3 Angel Wings to be refractions from a chamfer in the L3 glass.

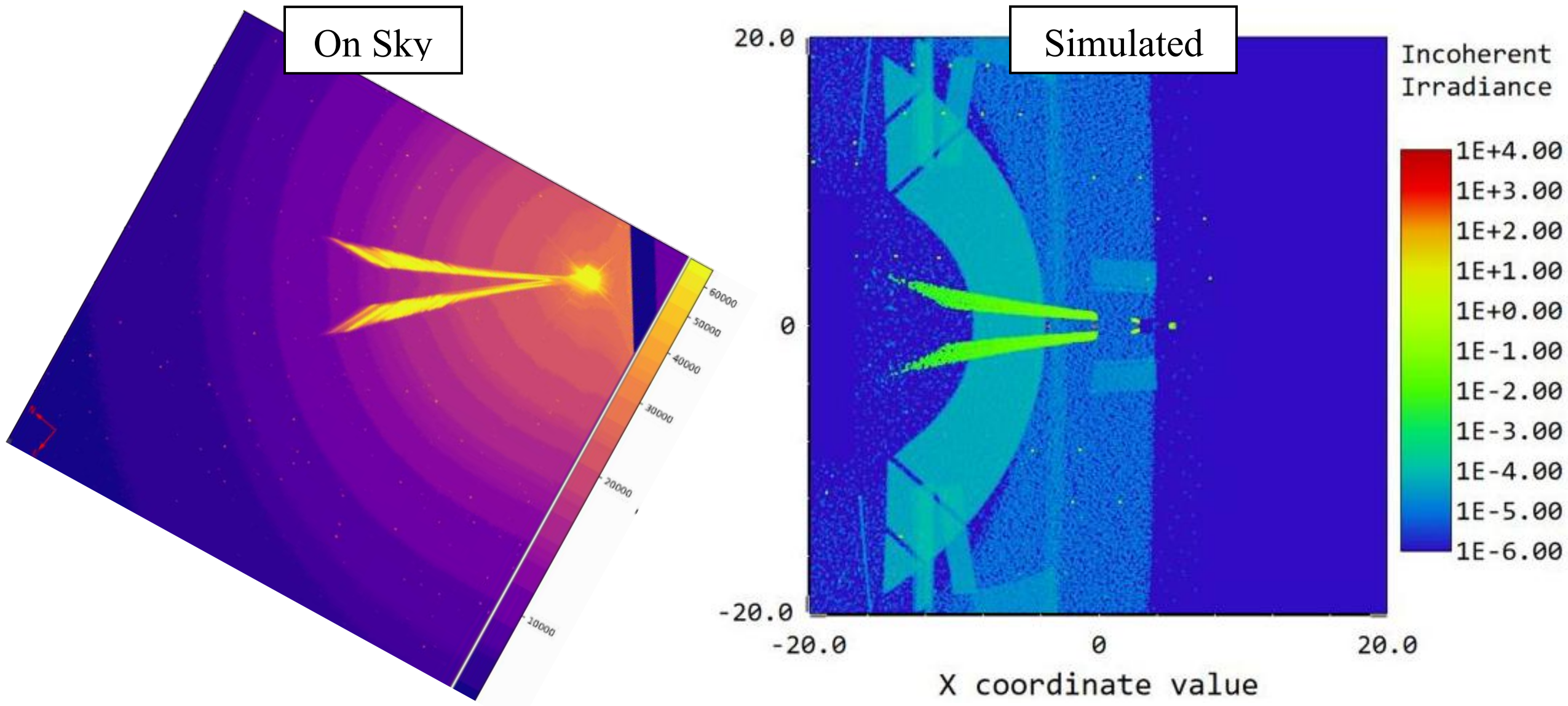

Figure 2: Another instance of photographed L3 Angel Wings at the left (visit = 2025051500395) versus the features produced by the Zemax model at the right show how similar the two look side-by-side [2, slide 8]. This simulation was used to perform the ray tracing in Zemax that then identified the source of this stray light artifact.

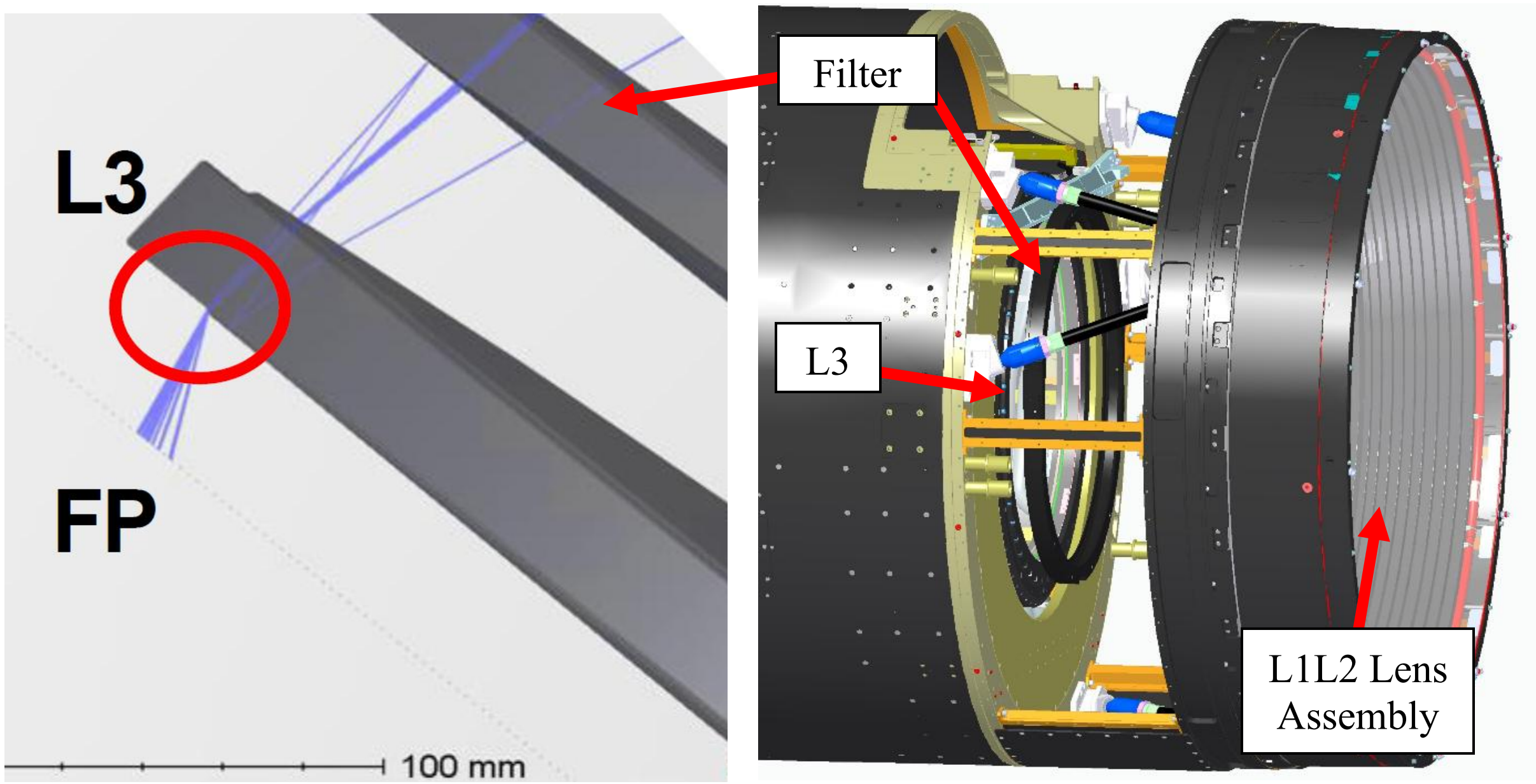

Figure 3: In the lefthand image, the region circled in red (where the rays converge) is the location of the L3 Chamfer [2, slide 8]. The ray tracing shown enabled the team to follow the rays from the stray light features back to originate at this location, proving it to be the cause of the L3 Angel Wings. At the right a stripped-down LSSTCam Assembly is shown to provide perspective/context regarding the location of the filter and L3 within the camera (many items are hidden to allow sight lines of this hardware).

The L3 angel wings bear significant resemblance to a stray light feature that was identified in images taken by the Dark Energy Camera on Cerro Tololo over a decade ago [3]. These features were subsequently mitigated through the efforts of the Tololo team, through a combination of baffles and antireflective paint that were added into the system as a result of a stray light mitigation effort similar to the one currently being carried out by the Rubin team [3].

The second variety of stray light artifact that will be touched on in this paper is named the “Horseshoe” due to the swooping U-shaped characteristics illustrated in Figures 4 and 5. It was initially noticed in the Zemax simulation in early June of 2025, then identified in an on sky image shortly after.

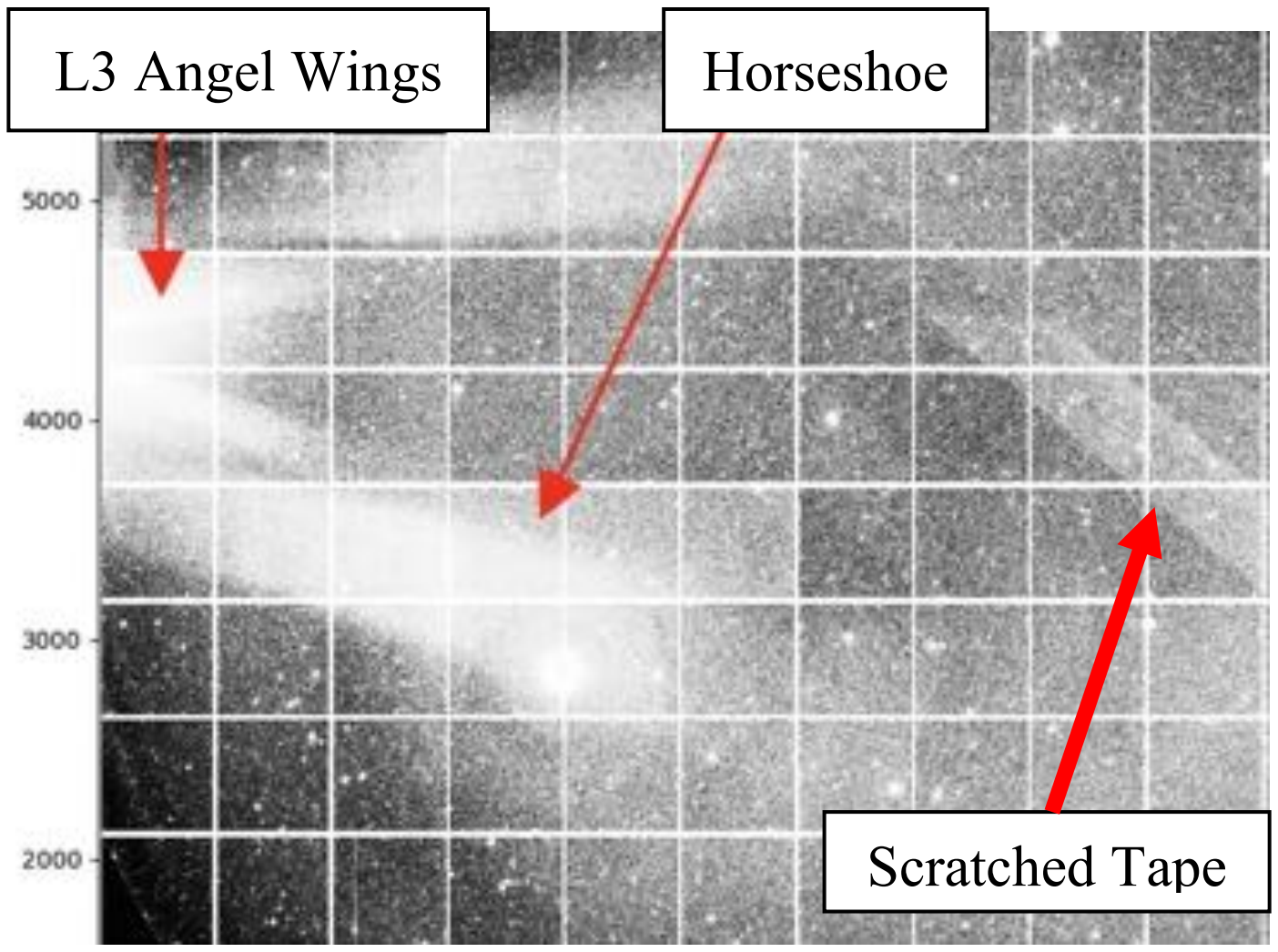

Figure 4: A large, fairly diffuse Horseshoe artifact can be seen spanning a large portion of the image above (less than half of a full focal plane image), with smaller Angel Wings also visible as indicated at the far left edge (visit = 2025111615439) [4, slide 7]. Also seen at the right side of this exposure is a different stray light feature, referred to as “Scratched Tape.” The Scratched Tape artifacts have been mitigated through the installation of a different baffle, which was designed and implemented thanks to the efforts of the Rubin stray light working group and is discussed in greater detail in Paper 14147-121 [5].

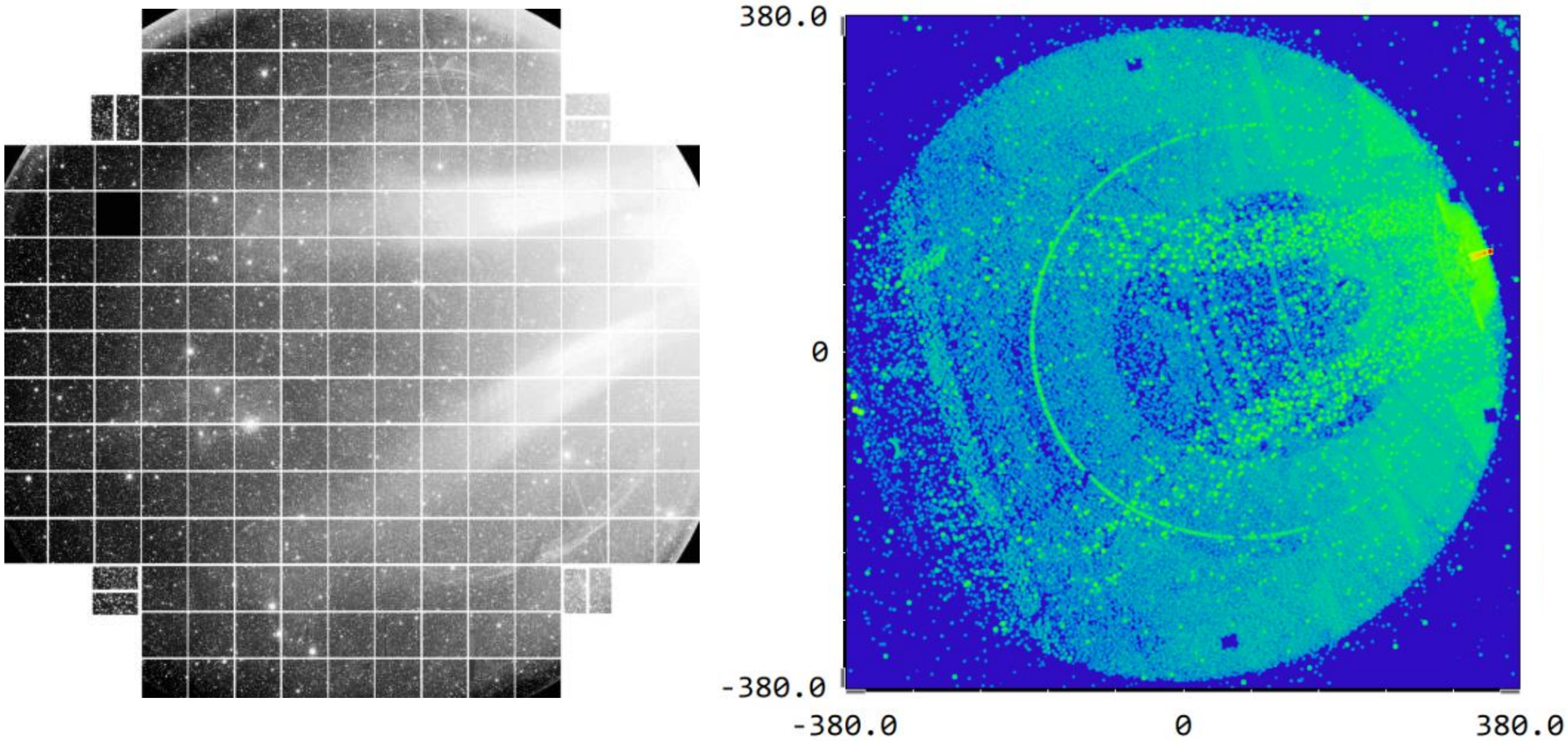

Figure 5: Another Horseshoe artifact (full focal plane exposure) at left (visit = 2025112637439), and a Zemax image of a simulated Horseshoe of near-identical morphology at the right. These simulations were key in being able to identify and characterize each artifact and trace them back to their origin point within the system.

The off-axis occurrence range for the Horseshoe artifacts is slightly narrower, at 1.994 degrees to 2.05 degrees. Note that this range is fully encompassed within that of the L3 Angel Wings – this is because these artifacts are ghosted reflections that originate from the same chamfer in the L3 glass. Refractions from the L3 chamfer are then reflected by the inner mechanical support of the L3 and ghosted from the L3 and/or filter glass back towards the focal plane, causing the larger, more diffuse effect.

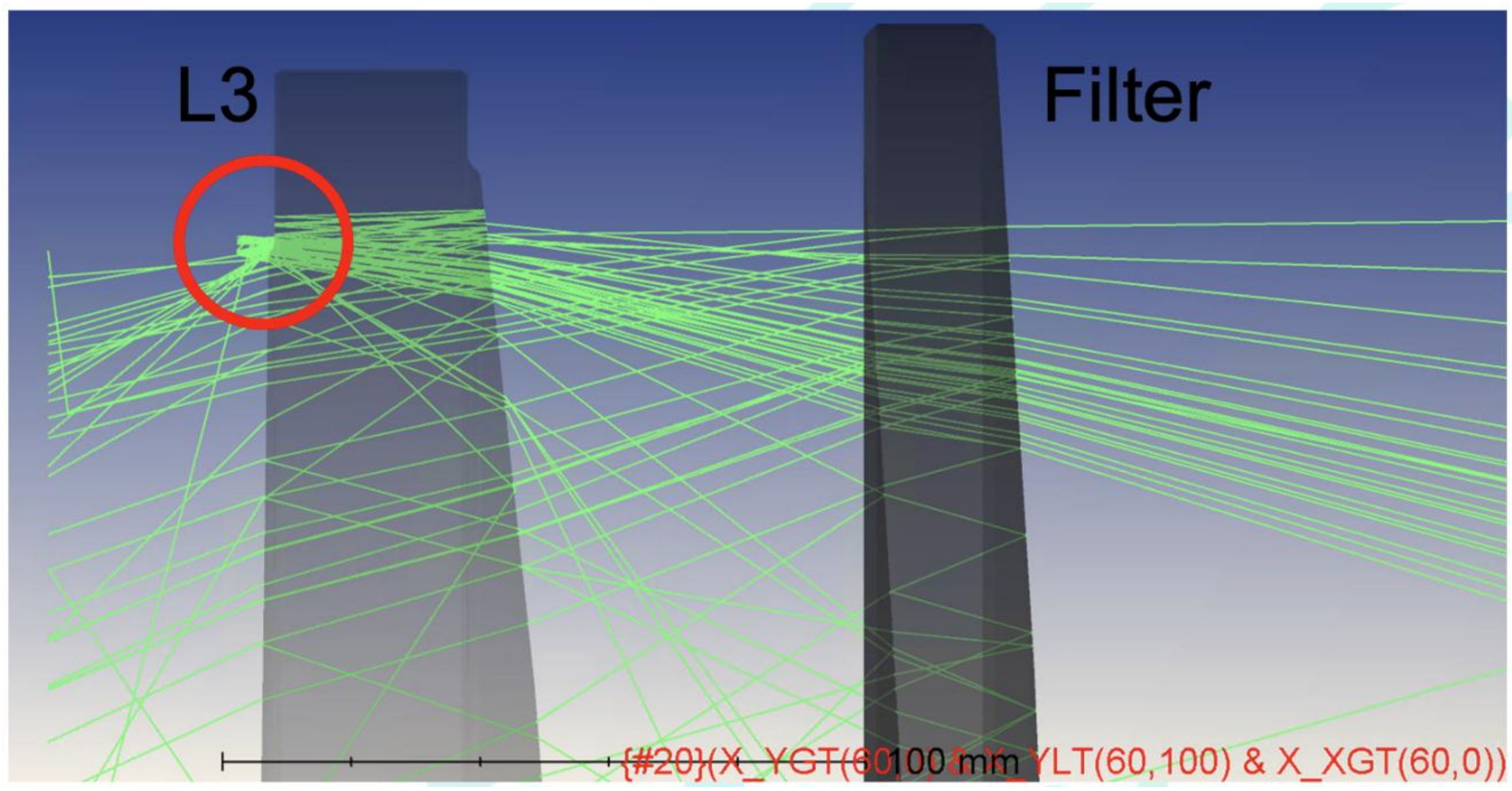

Figure 6: Similarly to the ray tracing performed for the L3 Angel Wings, a ray trace was performed working backwards from the Horseshoe features, which were found to originate from the same L3 Chamfer (circled in red) but after having ghosted from L3 and/or the filter, causing the more diffuse effect [4, slide 7]. The focal plane is located at the lefthand side of this figure where the rays all come to an end at one vertical line.

## 3. BAFFLE CONCEPT DEVELOPMENT

Once these stray light artifacts were characterized and traced back to this specific chamfer on the L3 the next steps were to determine the feasibility of adding a baffle to the already complete system. Not only did the team have to determine where it might be physically possible to install this baffle, but whether or not it would provide useful mitigation of the Angel Wings and Horseshoes without taking away from image quality in other ways (i.e. vignetting).

### 3.1 Location and Mounting Considerations

The first factor that had to be considered when investigations began was whether there was a location where the team would actually be able to install a baffle that would mitigate the problem with minimal impact on camera operations or the quality of the survey.

To avoid causing irreparable damage to the complete, functional LSSTCam assembly or its surrounding observatory hardware, it was quickly determined to be out of the question to drill and/or tap new holes in-situ. Thus, mounting options for installing a baffle were highly restricted, the team was limited to utilizing existing features within the camera. Fortunately, there happened to be an existing hole pattern of eight vacant, tapped holes on the front of the L3 frame that had previously been used during lens assembly and handling, and no longer served any purpose. These were immediately identified as the best possible mechanical interface for the theoretical baffle.

Not only did these holes provide a feasible way to mount a baffle in an area of the camera that would be able to potentially mitigate this stray light path, but despite this region being densely packed with hardware, there was sufficient room to install low-profile button head cap screws in seven of the eight holes to affix a baffle here. The baffle and mounting hardware, if installed, would be safely tucked away from the Shutter blades and the moving components of the Filter Exchange System, located behind a stationary part of the Shutter (called the garage plate) as seen in the lefthand image of Figure 7 below.

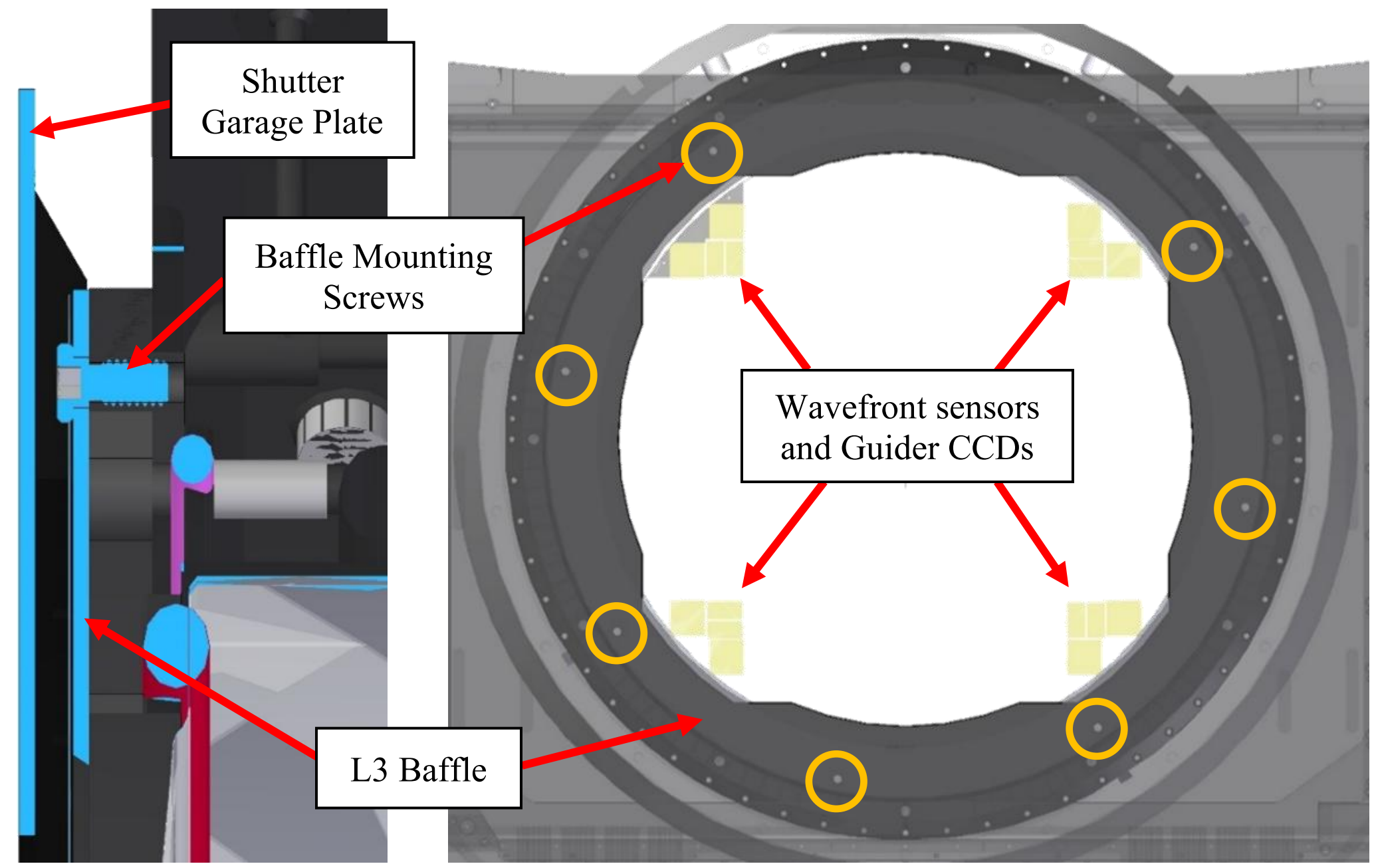

Figure 7: Selected low-profile mounting screws would have a minimum clearance of 2.8 mm when installed in existing hole pattern as seen in cutaway closeup (Left), and 7 of the 8 holes allow this clearance designated by orange circles (Right), which would be more than sufficient to secure the baffle to the L3 frame.

Once the potential mounting holes and hardware were identified, access needs in order to install this baffle into the camera were also considered. It was essential to be able to install this baffle (or remove it if needed) without needing to remove the LSSTCam from the Telescope Mount Assembly (TMA). The process of removing the LSSTCam from the TMA would entail extensive mechanical work and a significant amount of downtime from survey operation and was thus out of the question. The baffle and mounting screws can be installed while the camera is in-situ, mounted on the TMA, utilizing the deployable platforms and a handful of preparation steps to get other hardware out of the way first (including the Shutter and Auto Changer – part of the Filter Exchange System). See Figure 8 below where the TMA deployable platforms and camera are shown and refer to Section 4.5 to see further details about the time required to complete the preparation and installation of the baffle.

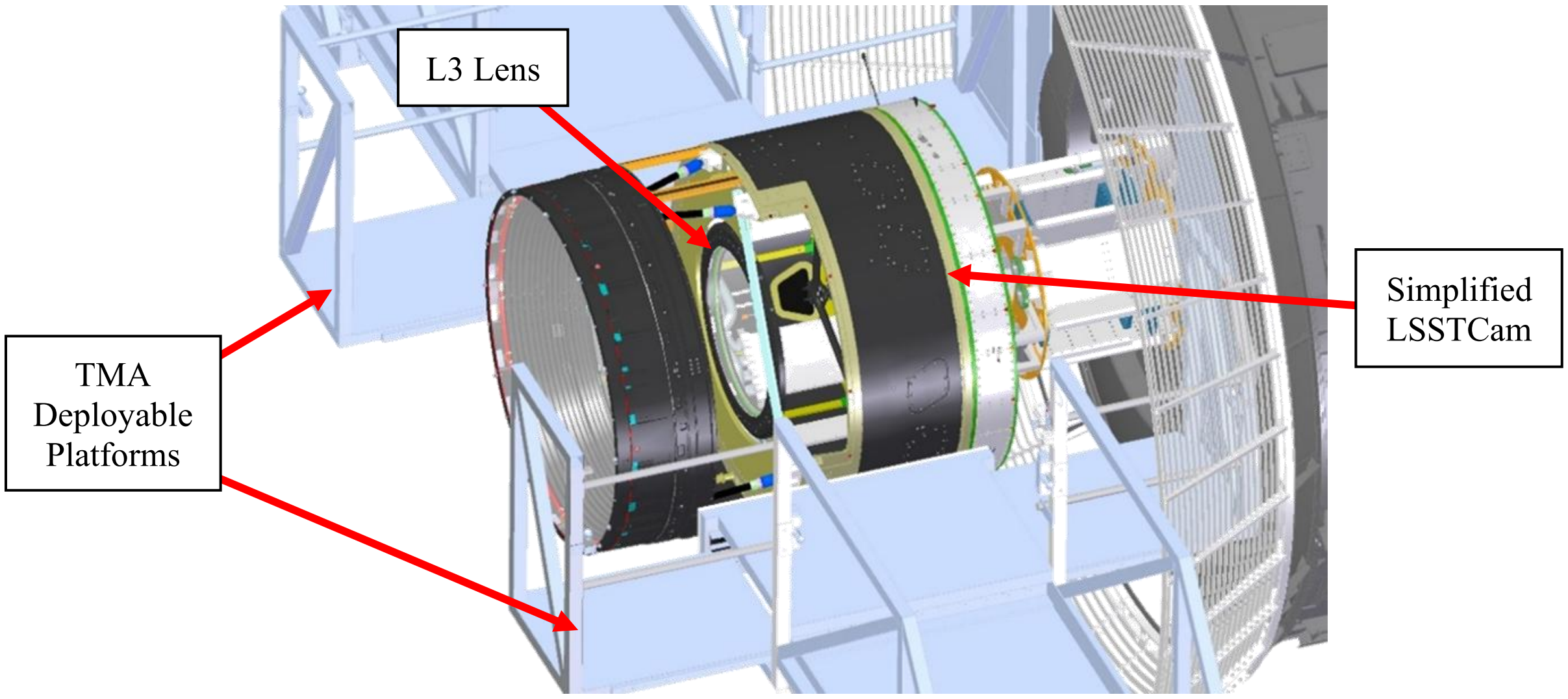

Figure 8: The LSSTCam is shown here mounted on the TMA (simplified, with parts not pertinent to this study hidden). In this image it is depicted at the 60 degree rotator angle used for shutter removal, with the platforms deployed on both sides for personnel access. Note that the Auto Changer and Shutter have been removed, as they need to be for this installation procedure to take place.

Aside from not impacting the two major mechanized systems in the region (Filter Exchange System & Shutter), it was also necessary to consider the other local infrastructure, which in this case is the airflow across the L3 front surface (L3-S1). Purge air flows out through the Shutter Garage Plate and across the non-vacuum side of the L3, preventing condensation from forming on the lens through warming the L3-S1 surface via filtered, dry, temperature-controlled air [6]. Additionally, this airflow contains minimal eddies with the goal of preventing the development of natural convection cells, which could affect image quality, and the constant flow also helps prevent particulates from alighting on the surface of the glass and sticking there [6]. Because the jets that inject this air into the area are located within the Shutter Garage Plate, an important consideration was whether the air circulation would remain sufficient if this baffle were to be installed in the region. Pertinent data from the original computational fluid dynamics (CFD) analysis can be seen summarized in Figures 9 through 11 below and given the importance that this purge system functions properly. This analysis was originally performed in 2017, so a new version is being created to repeat the analysis with the baffle added to quantify its impact. The new round of this CFD analysis is actively being developed and is discussed further in Section 4.3.

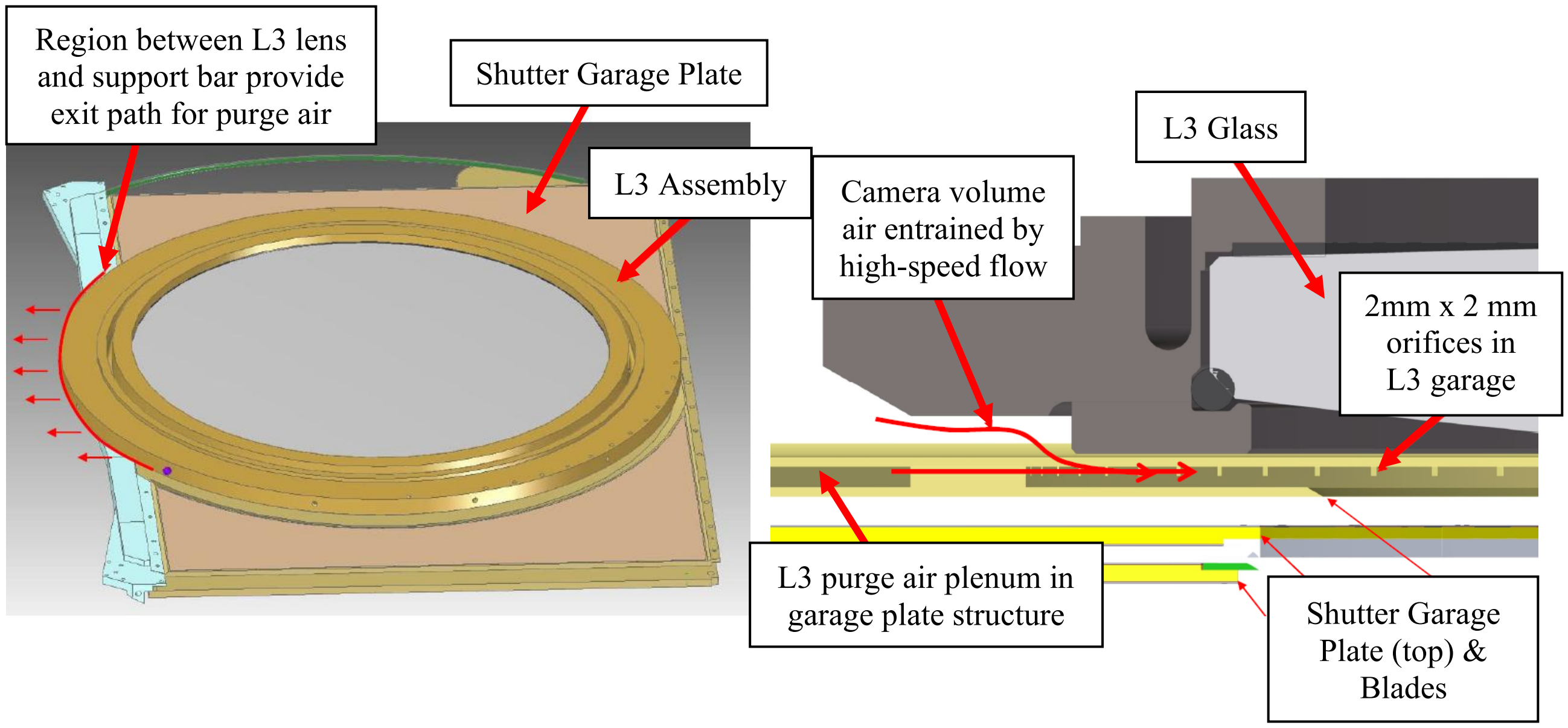

Figure 9: Overview of the airflow across the L3 region, showing shutter and L3 assembly in full (left), including the region at the +Y side of LSSTCam where the L3 protrudes beyond the top bar of the shutter and much of the purge air exits [6, slide 6]. Cutaway view (right) of shutter and L3 provides up-close airflow details and inlet orifices [6, slide 5].

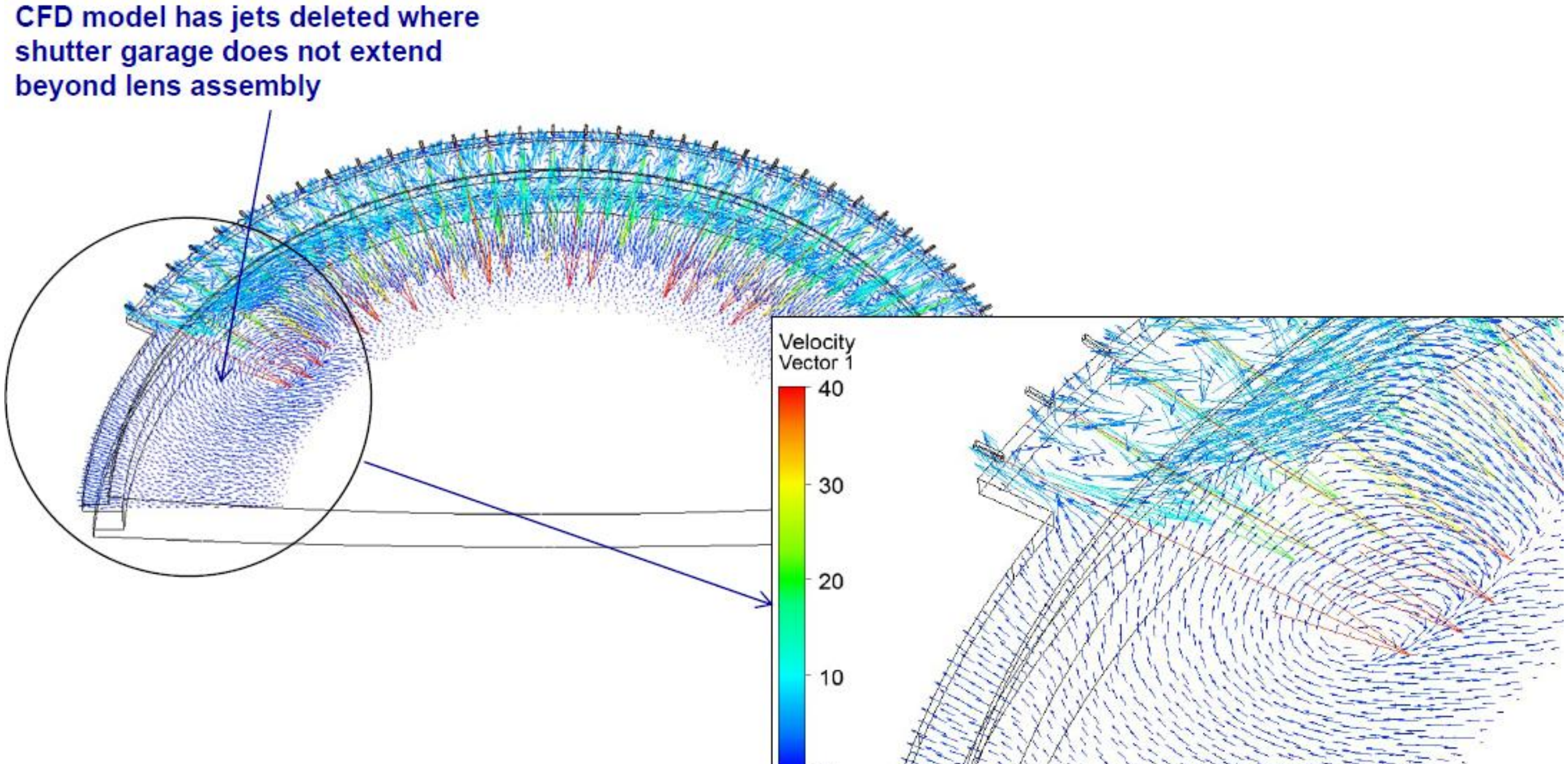

Figure 10: Results from 2017 CFD analysis in which air flow vectors are visible, with their velocities color-coordinated to the included scale in meters per second. Note that the exit creates an eddy, but that much of the airflow does still end up escaping through said exit. Additionally, the model is split in half with the YZ plane as the cutting plane as it is symmetric about that plane and this simplifies the model for running CFD more easily [6, slide 11]

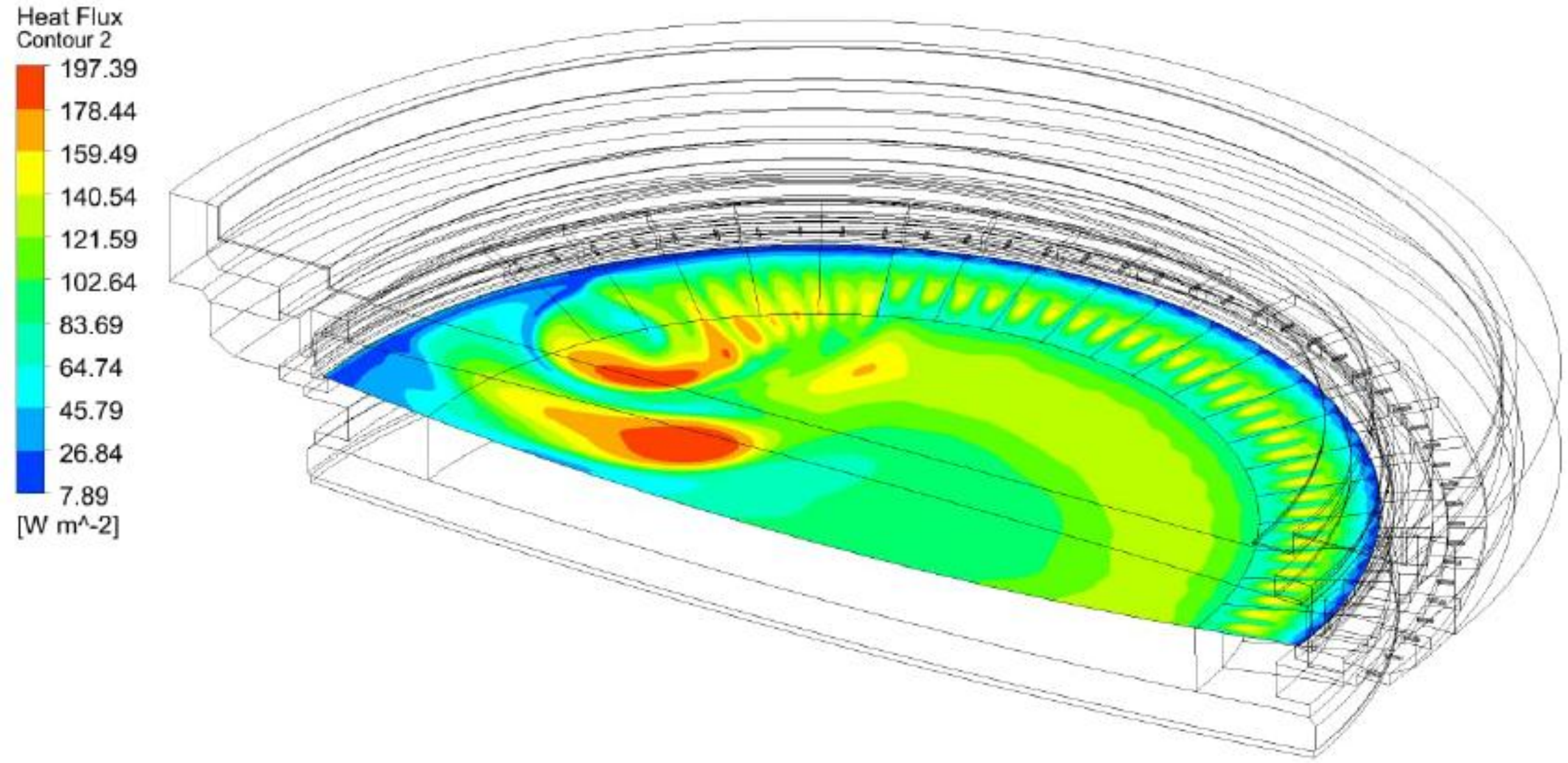

Figure 11: This figure displays the heat flux for L3-S1 from the 2017 CFD analysis, again utilizing the half-model with symmetry about the YZ plane of the LSSTCam. The heat flux contours mirror the air currents displayed in Figure 10, both from the inlet jets and from the eddy and exit region [6, slide 13]

In an ideal world, the mechanics of a stray light baffle would be guided by the optics of the instrument, to precisely block the light rays in question. However, in this case the need to utilize an existing mechanical interface combined with the tight space constraints of the L3 region forced the team to work at a given Z coordinate position. This constraint meant that the primary geometric aspect of the baffle that could be driven by the optical system was the size and shape of the aperture, as discussed below in Section 3.2 and in further detail in Section 4.1.

### 3.2 Preliminary Zemax OpticStudio Results

With the most viable mounting location identified, the next step was to test whether or not a baffle installed in this location could successfully mitigate the L3 angel wings and horseshoe stray light features. This analysis began with a simplified version of a ring-shaped baffle which was placed in the Zemax model at the mounting location identified in the previous section, allowing the team to determine the usefulness of installing a baffle in this location.

The physical diameter of the L3 glass is 782 mm (though part of this is obstructed by the L3 frame), the diameter of the S2 chamfer (the source of these artifacts) at approximately 729 mm, and the L3 clear aperture has a minimum diameter of 722 mm. It was also known that the off-axis range of the L3 angel wings was 1.99° to 2.055° (with the range for the horseshoes contained within this range, at 1.994° to 2.05°), and from these various known values, the initial optical study for this proposed baffle utilized an aperture diameter of 680 mm (1.985° off-axis angle) at the mounting location on front frame of the L3 assembly.

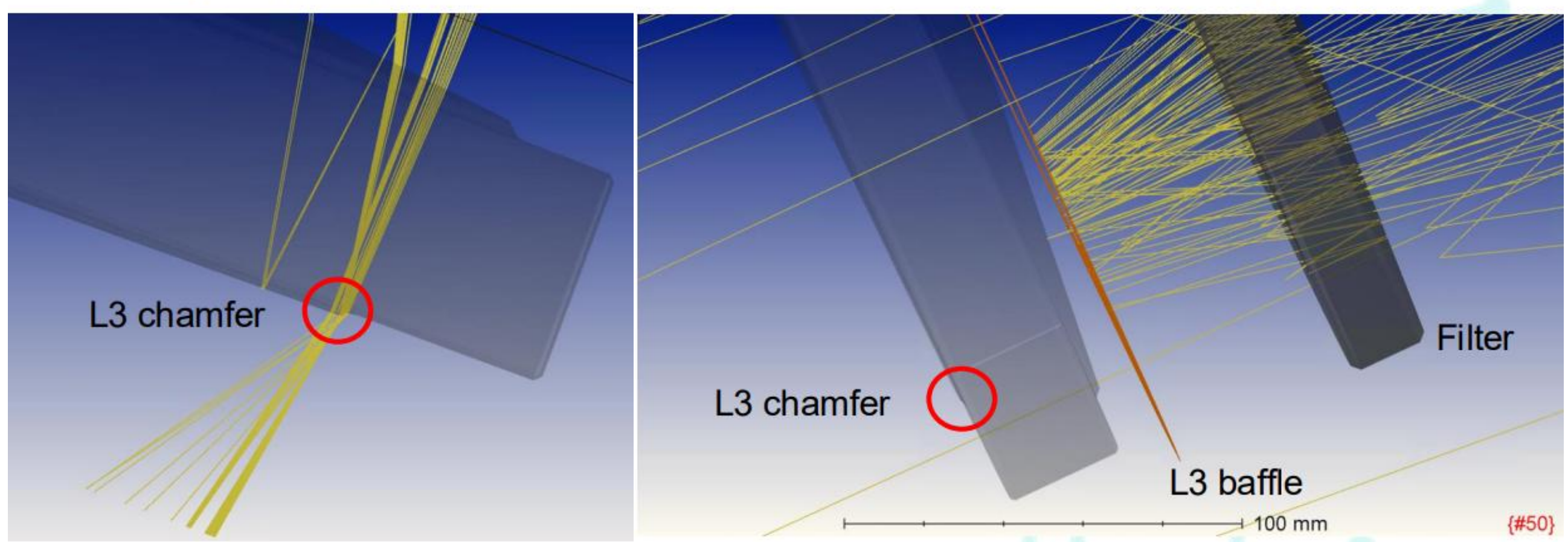

Figure 12: At the left the current LSSTCam configuration is displayed, where light rays of the applicable off-axis angle range (1.99° to 2.055°) can be seen hitting the L3 chamfer and reflecting/refracting to create the L3 angel wings and horseshoes. On the right side the conceptual baffle design with a circular aperture of 680 mm has been introduced into the simulation (displayed in orange), and it prevents a large portion of the offending light rays from reaching the focal plane [4]

This baffle proved successful in blocking the rays that reflect/refract off of the L3 chamfer, thus mitigating both the L3 angel wings and horseshoes in the Zemax simulated images that were produced. The images in Figure 13 were selected to summarize the results of this first Zemax simulation; they display the difference between stray light distribution and morphology with and without the baffle in the model at three different off-axis angles.

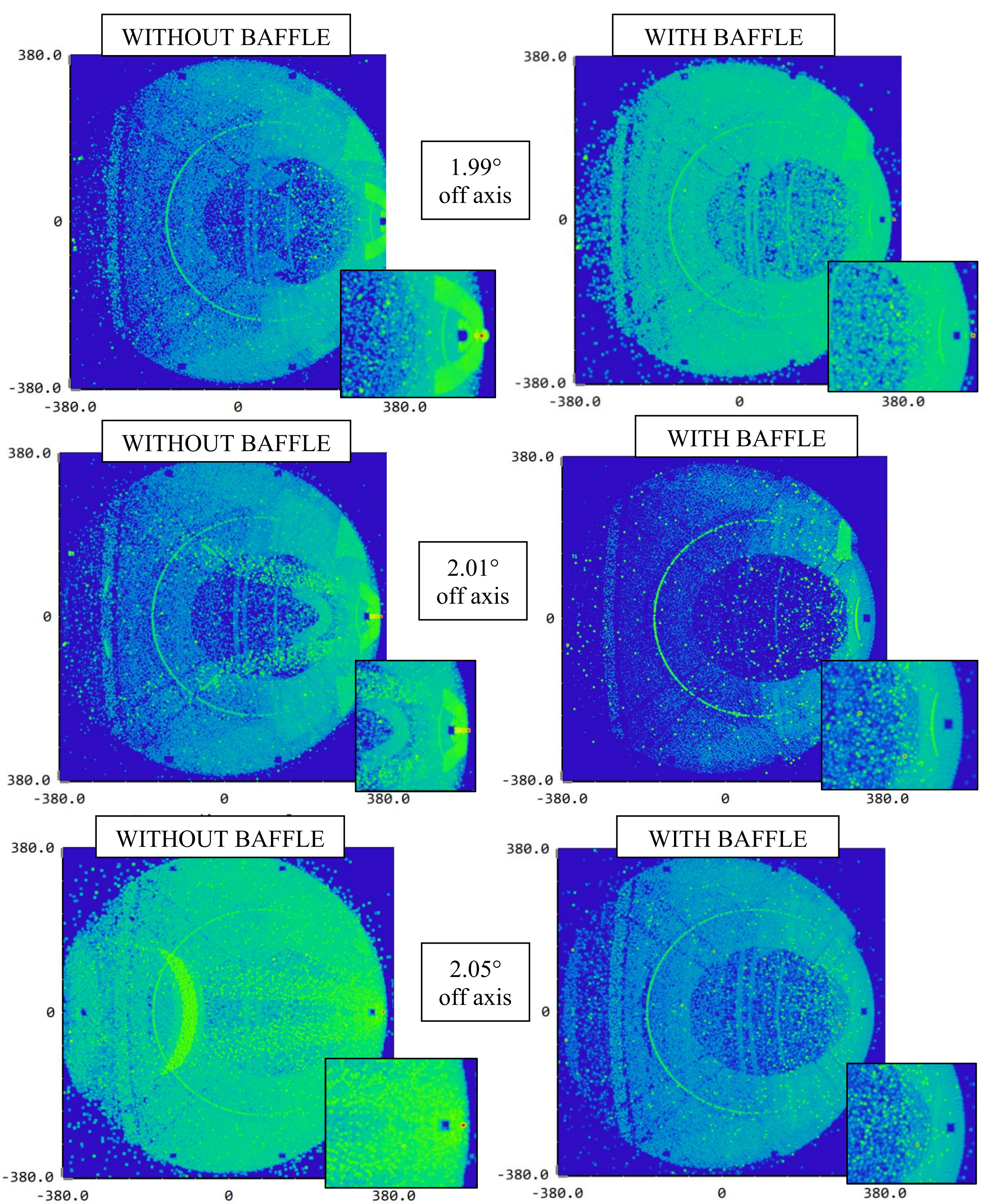

Figure 13: The images in this figure show the distribution and intensity of the two features being studied, both without and with the baffle added into the simulation. This process was repeated for multiple off-axis angles (labeled accordingly), and intensity of both the horseshoes and L3 angel wings are noticeably mitigated by the baffle. Note that these results are from the final baffle design, but earlier iterations provided similar results, proving the concept useful.

These simulation results proved very promising regarding the potential stray light mitigation that an L3 baffle could provide, showing a significant reduction in both the L3 angel wings and horseshoe features. With this serving as a proof-of-concept, the group continued with more in-depth development of the proposed L3 baffle, however significant amounts of further iteration and analysis were needed in order to hone the design in to be as beneficial as possible to the Rubin Observatory – this will be covered in detail in the following section.

# 4. BAFFLE DESIGN

After determining that it would be possible to physically mount an additional baffle onto the L3 frame, then confirming that it had the capacity to mitigate the stray light features in question, the next focus was to converge on the most ideal baffle design. Different aspects (mechanical, optical, integration, and fluid dynamics impacts) had to be considered, driving design updates which then required verification. This iterative process narrowed down a baffle geometry that would be the most beneficial to the survey while also not interfering with system performance in other ways.

## 4.1 Aperture

Building upon the preliminary prototype that was covered in Section 3, the next step was to iteratively update the aperture sizing and geometry to prevent vignetting while still blocking as much stray light as possible.

At the outermost corners of the LSSTCam's focal plane there are four "corner rafts" which are sensor packages similar to those that make up the rest of the focal plane, except that instead of 9 charge-coupled devices (CCDs) for imaging, they each have two "guider" CCDs and two smaller intra- and extra-focal "wavefront" sensors [7]. As can be seen in the figure at the left below, this preliminary concept (fuchsia) visibly blocks a portion of the guider sensors and once taking into account the angle at which light enters the L3, the baffle was also likely vignetting the wavefront sensors. To cut down on this vignetting, the next prototype design of the baffle had cutouts at each corner (seen at right below), which were offset to be slightly farther out than the outer edges of the wavefront and guider sensors.

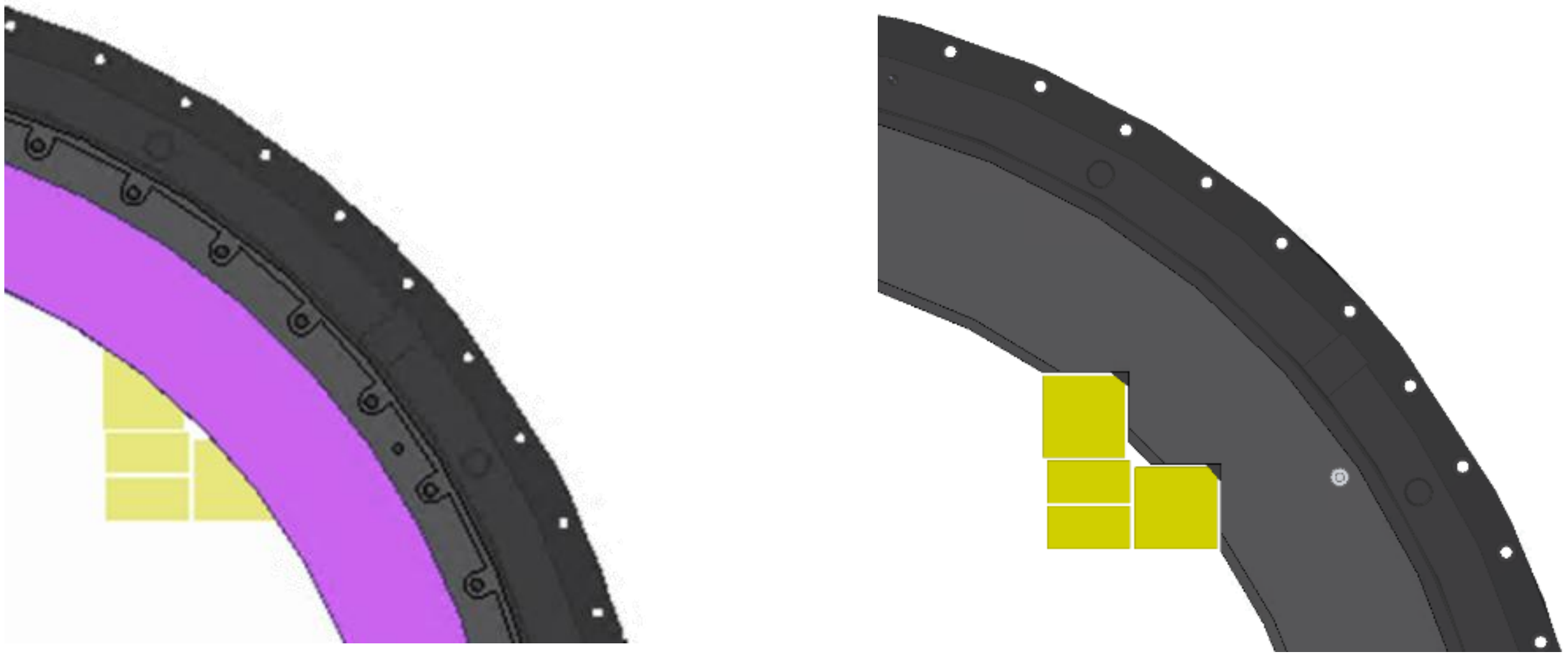

Figure 14: The preliminary conceptual baffle discussed in Section 3.2 is shown in fuchsia at the left. It has a circular aperture of 680 mm but can clearly be seen covering a portion of the guider sensors in the image. To attempt to remedy this, the next version of the baffle design (at right) added a stair-step cutout to clear the sensors of the corner rafts. Note that only one corner is shown in this figure, but the vignetting affected all four corner rafts similarly

A new Zemax simulation was performed using this modified baffle prototype, and this time an assessment was performed to investigate the vignetting of the guiders and wavefront sensors when compared to the current system. It found that despite the added corner cutouts, there was still a vignetting factor exceeding 15% for the wavefront sensors' fields [8]. Below are the graphical results from this investigation, which make it evident that the cutouts in the corners *do not* meet the goal of not inducing any additional vignetting at the wavefront and guider sensors.

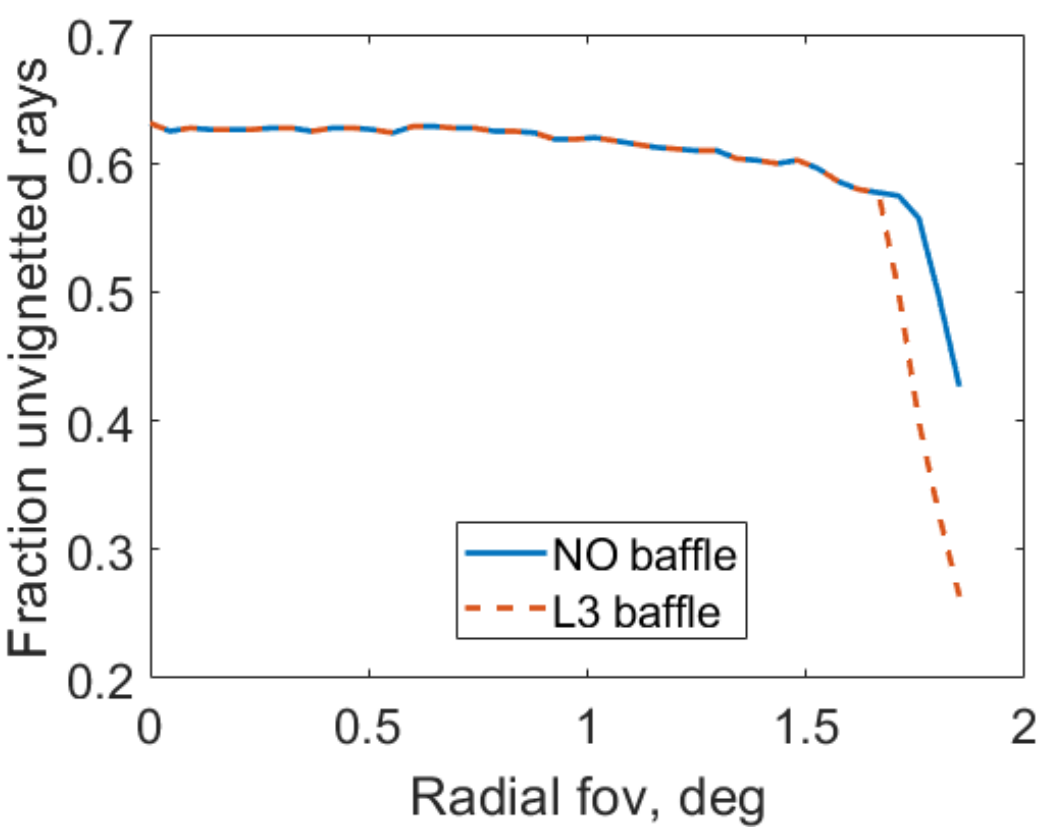

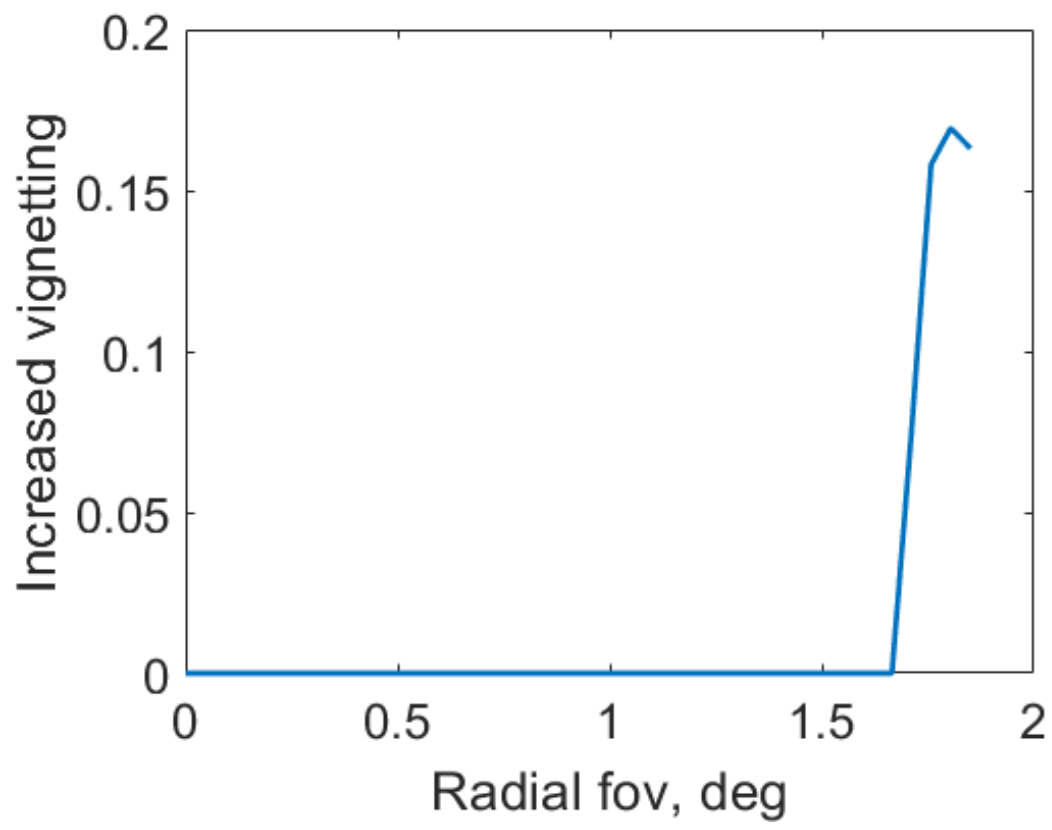

Figure 15: Using the updated prototype (with stair-step corner cutouts), additional simulations were run to see if vignetting was a concern with the new cutouts. A s can be seen in these two graphs, even with the cutouts, the baffle caused about 15% additional vignetting of the sensors on the corner rafts (wavefront and guider sensors), requiring a redesign of the corner cutouts [8]

The stray light working group then continued the iterative process utilizing Zemax ray tracing to reverse engineer a new version of the baffle that would not add any vignetting effects. Ray trace geometry which contained the optical elements of the LSSTCam as well as approximately 3000 rays converging onto one of four corner rafts (Figure 16) was exported from Zemax and imported into Solid Works, where we mapped where the rays intersected the Z position of the proposed baffle. This map of intersection points served as a stay-clear area for this corner, as seen in blue in Figure 16. This was deemed to be the most accurate way for us to hone in the baffle cutouts without having to repeatedly iterate based on conjecture and approximate light entry angles alone.

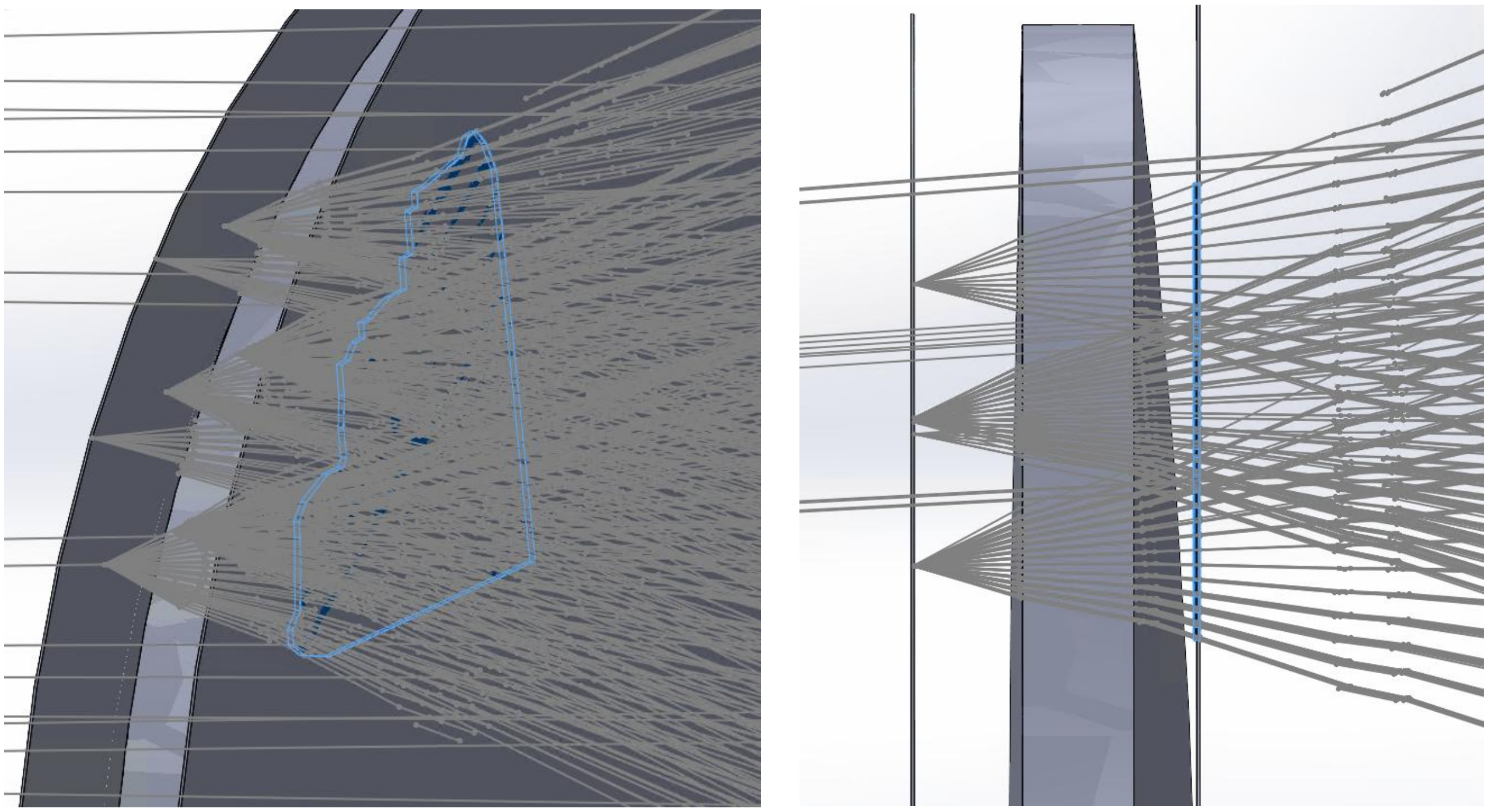

Figure 16: Ray tracing was utilized to reverse-engineer the corner cutouts on the next iteration of the L3 baffle, ensuring that no added vignetting effect would be introduced by the baffle. The rays were used to create the new required clearance for the corner cutout geometry. Only one corner is shown, but the symmetry of the system allows for this same geometry to be mirrored to the other corners.

We then brought this stayclear area into the larger camera assembly model, and mirrored it about both the XZ and YZ planes to find the stayclears for all four corner rafts (this was deemed a fair assumption due to the symmetry within the optical system and the fact that the L3 chamfer exists around the entire circumference of the L3 glass). Once these stayclears were in place, we modified the corner cutouts to create a new baffle prototype, shown in Figure 17 below. The inner radius maintains the same clear aperture radius as the initial prototype (680 mm) to block stray light wherever possible, while the corner cutouts are designed to clear the geometry derived from the ray trace study. This new prototype was dubbed the "Mickey Mouse Baffle" due to the modified geometry resembling his famous ears – it is referred to as such in this paper and the figures therein.

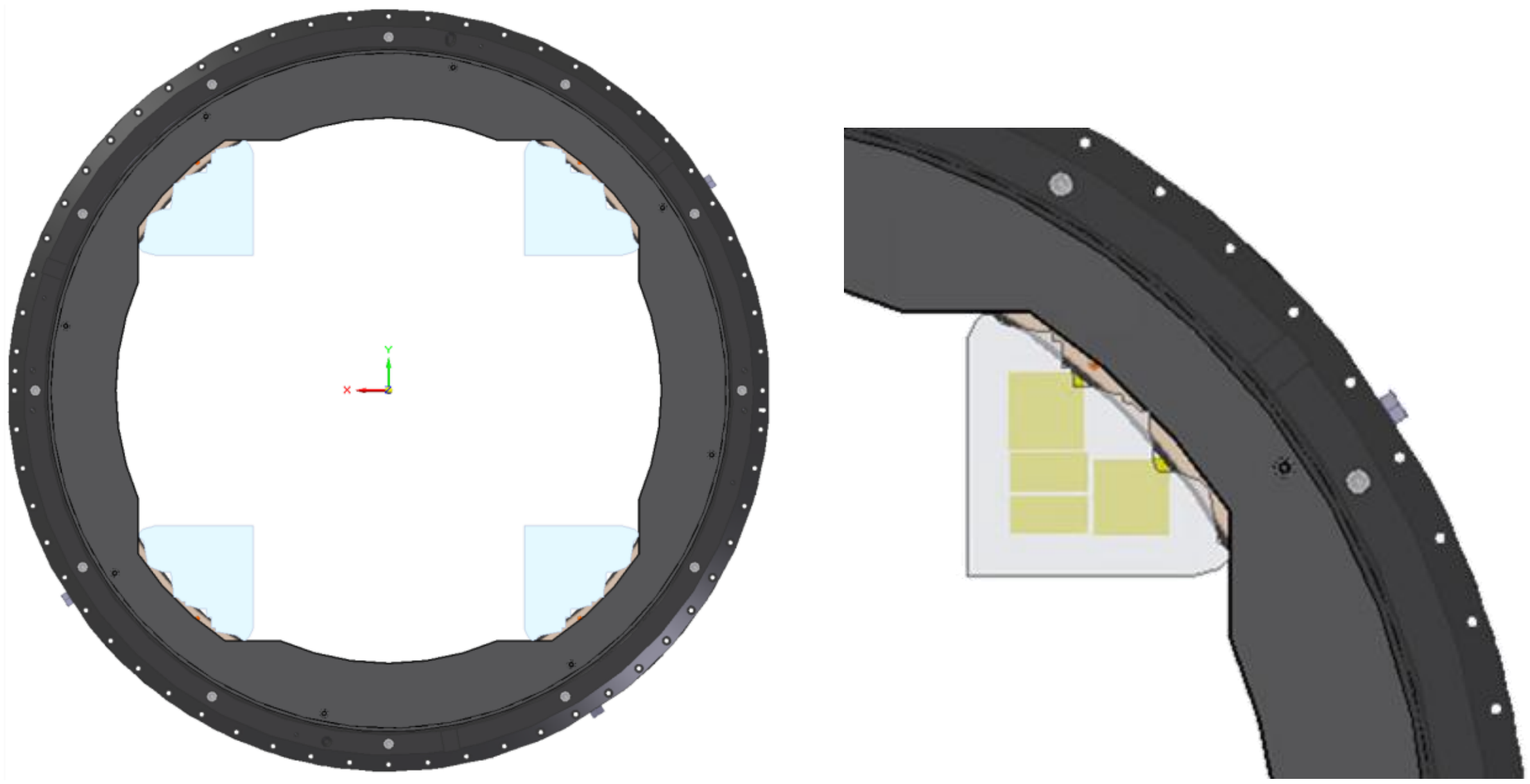

Figure 17: The "Mickey Mouse Baffle" geometry is seen here, in full at the left side, where the grey shapes in the corner are the stay clear areas determined by the ray tracing as shown in Figure 16. On the right is a closeup of the +X+Y corner, with the grey stay clear made transparent in order to also show the corner raft sensors for reference

Preliminary Zemax studies of the Mickey Mouse Baffle proved much better at not vignetting the scientific field of view, wavefront, or guider sensors. The graph at the left in Figure 18 shows the vignetting without a baffle installed (current actual system, blue line) overlaid with this iteration of the baffle installed (red dotted line). The perfect alignment of these two curves (supplemented by the graph at the right below) display that the Mickey Mouse Baffle is expected not to add any vignetting to the system. When these results are contrasted against the graphs from the previous baffle iteration (Figure 15), we see the comparative improvement, but at the same time we can expect less blockage of stray light with the larger cutouts. There had to be compromise between stray light mitigation and additional vignetting, and the decision was made that it is most beneficial to the survey to mitigate stray light as much as possible while not adding any vignetting to the system. Especially now that these stray light features have been identified, the team is able to identify them, flagging and removing any affected areas from coadded or stacked images, though the less images that get excluded from such images, the better.

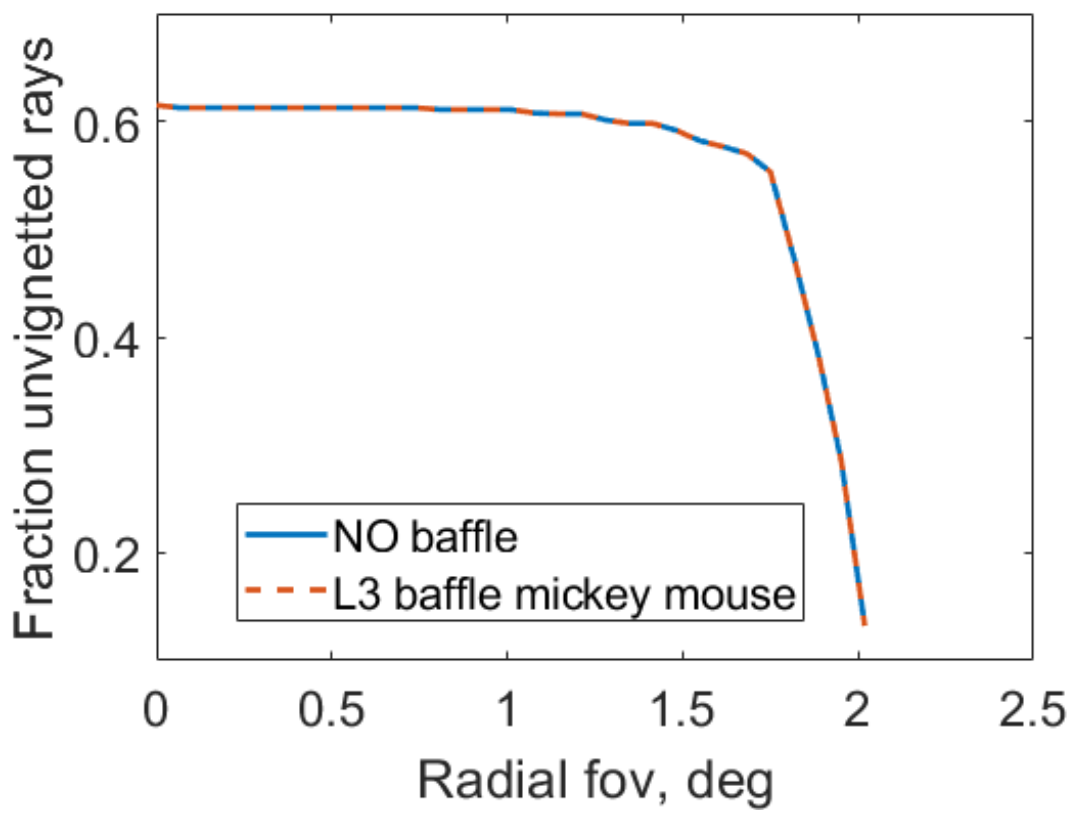

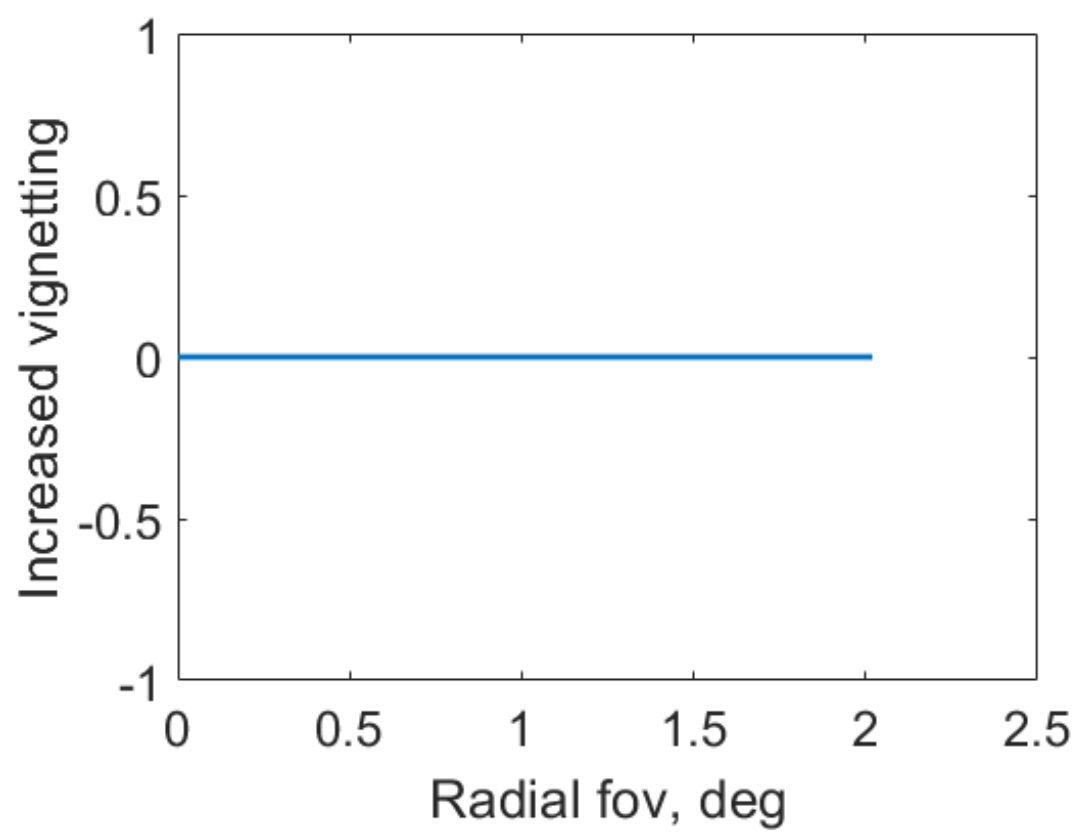

Figure 18: With the updated "Mickey Mouse" baffle aperture, the amount of vignetting remained identical to the system with no baffle. This is displayed by the visible overlap of the lines in the graph on the left, then verified by the right-hand graph showing no increase in vignetting throughout the entire field of view [9]

### 4.2 Edge Bevel

Complimentary to the geometry of the aperture cutout, the edge of the inner opening must also be considered to provide the best baffling possible without adding new surfaces that have the capacity to introduce additional stray light features. Knife-edges or bevels are customary in light baffles, but it is vital that their orientations and angles be designed to prevent them from being illuminated and critical simultaneously, as to not introduce new low-order reflections while attempting to mitigate other artifacts [10]. Since this is one solitary baffle as opposed to the numerous vanes internal to a cylindrical baffle as covered in Fest's calculations, the formulas for deriving specific edge angles do not perfectly apply, however, the importance of not being both critical and illuminated remains [10].

Due to traditional manufacturing constraints when machining on a standard or computer numerical control (CNC) mill, a 45 degree angle is often used for bevel edge geometry, but other options ranging from 15 to 80 degrees are available [11]. Manufacturing such an edge does require that the part is backed appropriately during the work to prevent the creation of a burr, maintaining the desired, sharp edge. Alternatively, more novel manufacturing techniques such as wire electrical discharge machining (EDM) can be considered and have the capability to create even more customized angles and sharp edges without the same burring concerns as traditional machining, but this method does require that the material be conductive [12]. The dilemma as the bevel edge becomes sharper is that the thin material becomes more at-risk to damage, which would negate some of the positive aspects, potentially causing strange reflections.

Given these constraints and the location of the baffle within the LSSTCam, the current prototype uses a 30 degree bevel angle. This geometry can be iterated in Zemax and honed as needed after all other simulation and analysis work is complete (vignetting and fluid analysis), assuming that the project-level decision is made to move forward with the production of this L3 baffle.

### 4.3 Fluid Analysis

In Section 3.1 there was a brief discussion of the purge airflow that goes across the L3 lens in this region of the LSSTCam, serving the primary purpose of preventing condensation from forming. Adding a new part into this area thus requires the team to analyze the impact it may have on that airflow.

The L3 purge air originates from 80 2 mm by 2 mm gas jets, each one injecting 0.125 grams per second of air at 0.7 atm and 5 °C below ambient onto the surface of L3. Initial purge airflow analysis for this area confirmed this purge system to adequately provide convection warming as well as conditioning of air for L3 by way of this constant airflow across the glass [6]. Should this airflow fail to serve its purpose, there would be major impacts on the ability of the LSSTCam to take quality images, and should condensation form, the need for lens cleaning could arise. Because of this, the stray light working group is working on conducting a similar simulation of the airflow in the area both with and without the baffle installed to see its effect on the temperature and flow of air across the lens.

The initial simulation was performed nearly 10 years ago and both the engineers who worked on it and the files themselves are no longer available to be consulted, so we began the process of recreating the analysis in Ansys Fluent. This began with the slight simplification of the hardware in the region: the Shutter, L3, Cryostat, Filter, and L3 Baffle. This simplified assembly was then brought into the CFD software (Ansys Fluent) where the air volume was extracted and the inlets, outlets, and part walls were identified. Conveniently, due to the symmetry of both the source airflow and hardware in the region about the Y-Z plane of the camera, the full problem can be set up with a half-model and the utilization of that symmetry. The remainder of this simulation is currently still under development, but the final results will be used to confirm that adding this baffle does not have adverse effects on the thermal conditioning of the L3 glass.

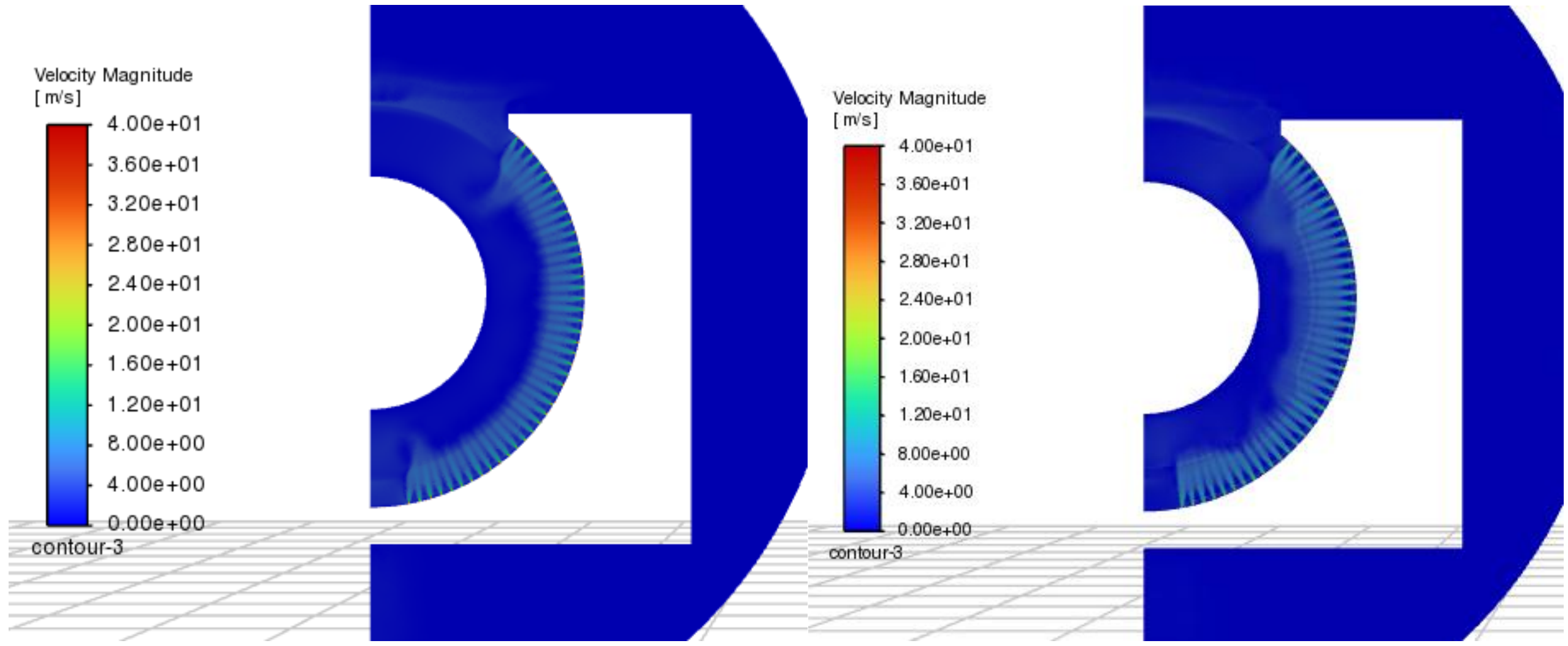

Figure 19: Shown here are images from the new CFD analysis which is currently under development. The contours of the air velocity exiting the Shutter Garage Plate (contour plane bisects the jets, simulation is symmetric about the center plane). At left is the current LSSTCam configuration (no baffle), and at right is with the Mickey Mouse Baffle, which can be seen shaping the airflow. Further CFD analysis of the system is underway to compare the temperatures of hardware in the region after the addition of the baffle. Both the 2017 and new version of the CFD have the shutter blades open.

### 4.4 Manufacturability

Promising simulation results can only get a hardware system to a certain point, but throughout the design process manufacturability must also be considered to ensure that execution of the system is possible. The topics of part geometry and materials are co-dependent, driving one another in different ways, meaning that their considerations had to be integrated into the design process to ensure that this L3 baffle would be able to be made and installed.

Geometric details of the baffle aperture have already been discussed in this paper, however there are other geometric aspects that carried weight in getting the baffle design to where it is today. Size is one of the first things to consider when having a part made, and this specific part's size limitations were driven by the location constraints. Mounting to the L3 front flange determined that the baffle had to be at least 33 inches (838 mm) in diameter to accommodate the mounting holes, but not much larger in order to not excessively overhang the L3 front frame. The maximum thickness was limited by the need to fit behind the Shutter, a gap that is approximately 0.22 in. (5.5 mm) at its narrowest, while the minimum thickness was driven by material and machinability. The thinner that a baffle like this is, the less likely it is to create reflections off of the inner edge, but this can be supplemented by the knife edge bevel discussed in Section 4.2, so the baffle was set at .079 in. (2 mm) thick to bridge the gap between sufficient clearance, performance, and manufacturability.

These size limitations are not producing a particularly miniscule or extremely large part, so the door is left open as far as material and manufacturing processes at this stage, but next to consider are the shapes that need to be achieved. The shape of the "Mickey Mouse" baffle consists of a circular outer diameter with bolt holes and an aperture cutout of arched segments and straight lines, as dimensioned in Figure 19 below. These features don't impart notable manufacturing limitations except for perhaps precluding fiberglass or carbon fiber construction (or at least making them quite difficult), but these manufacturing methods are ruled out for numerous other reasons and are thus not problematic to exclude. The final geometric consideration is that of the bevel edge, which was covered in detail in Section 4.2, and as mentioned there, remains open for iteration/discussion if the project moves forward with baffle production. Those iterations shall involve both discussion with fabricators regarding feasibility, and re-running Zemax simulations to confirm the final bevel angle.

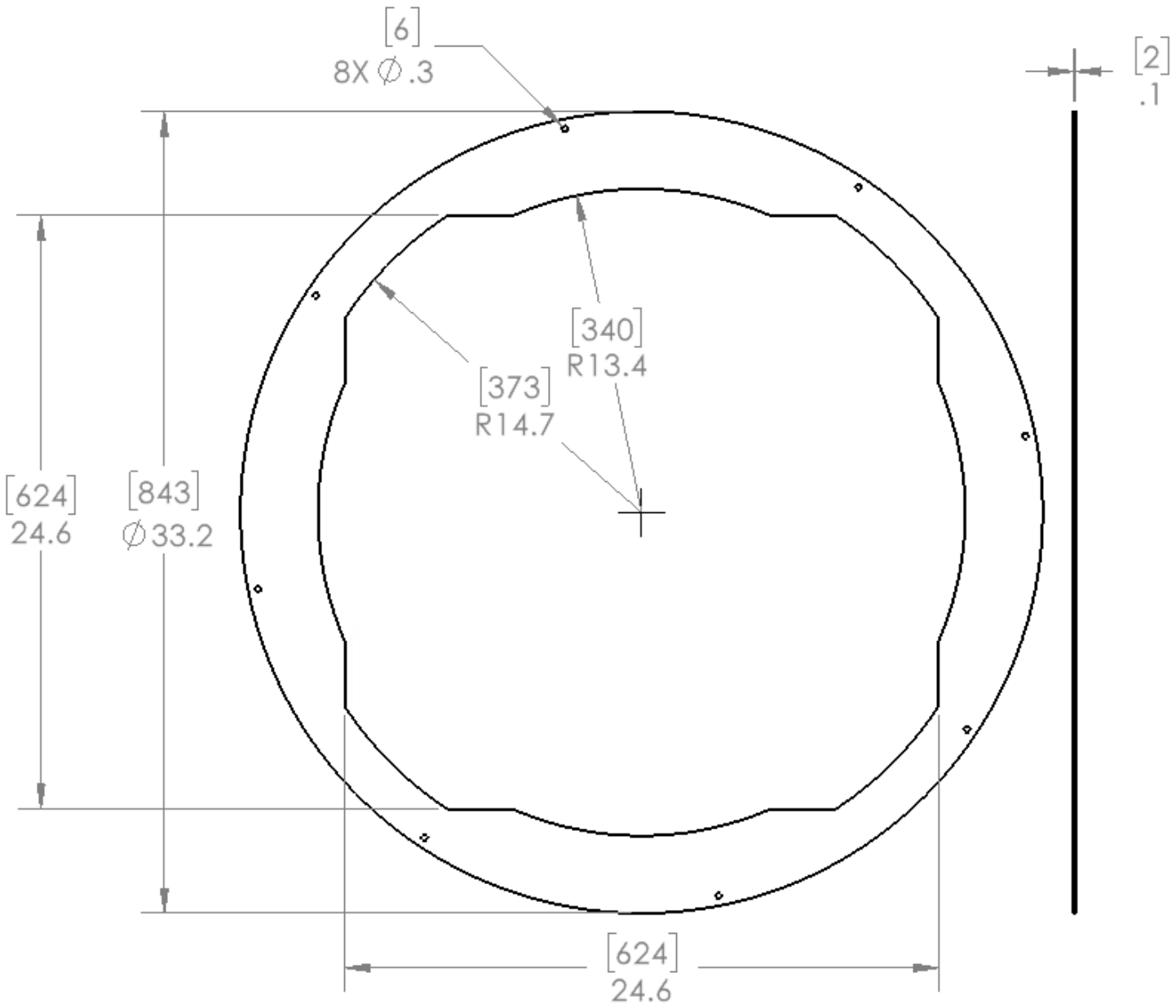

Figure 20: In this figure the overarching dimensions of the proposed "Mickey Mouse" baffle can be seen to give an idea of the scale for the sake of manufacturing considerations.

The decision of materials must be made in parallel with the geometric and manufacturing requirements and limitations, both that of the base material and the coating to achieve the best optical attributes. For a light baffle, especially one that would be mounted close to the focal plane, the team would like the blackest, flattest finish possible, but other aspects must be weighed to ensure that the part will hold up in the environment, being mounted on the Simonyi Survey Telescope long-term. The finish must be hard enough not to chip off when bolted into place, and must withstand a decade of use, additionally it must not outgas or shed particulates that would then get blown directly towards the L3 lens by the purge air. It must also be compatible with the material that the baffle is to be fabricated from.

The LSSTCam and the greater Rubin project have utilized Aeroglaze® Z306, Acktar, and Pioneer Metal Finishing finishes on parts in the past, and at SPIE Photonics West the team learned about Musou Black. The pros and cons of each of these options for the L3 Baffle Application are discussed below.

Musou Black, while it boasts impressive absorption rates of up to 99.4% for the visible spectrum, is an acrylic paint, causing concerns over the hardness and durability of the finish [13]. Pioneer Metal Finishing's Optical Black® finish boasts low outgassing alongside absorption of light from infrared to ultraviolet but lacks quantitative data and is only usable on Aluminum Alloys [14]. Acktar has multiple products to consider for this application, with the most promising of them being Vacuum Black™. It is low outgassing, works on all metals, is clean-room grade, and is only a few microns thick, making it a promising selection. However, Vacuum Black™ is applied through thin-film disposition, and due to the size of the baffle the team would have to consult directly with Acktar about the possibility of doing a custom production run [15]. Aeroglaze® Z306 similarly is low outgassing, compatible with any unfinished metals, and when applied with the correct preparation and curing processes provides a durable finish [16]. It has significant heritage both in the astronomical community and in the Rubin project itself, in fact the stray light working group has vendor contacts who are familiar with the priming, painting, and curing processes, and who have recently applied this finish to other large parts for the project. This confluence of factors results in Aeroglaze® Z306 as the best selection for the finish of this baffle. It is worth noting that Aeroglaze® Z306 was the same antireflective solution that was utilized to successfully mitigate the features that closely resembled the L3 angel wings in the Dark Energy Camera, giving it even more credibility as a viable option for the Rubin team's needs [3].

Finally, we have the consideration of delivery of the baffle to the observatory site, Cerro Pachon in Chile. Our two options are to either have the part completely fabricated and coated in Chile, or to make and coat the baffle in the US and then ship it down to Chile safely. The stray light team has connections in the San Francisco Bay area in California as well as Tucson, Arizona related to making all sorts of hardware for the telescope, including numerous parts that have optical black coatings of different varieties. This, combined with the fact that the mechanical engineer working on this baffle is located in California drive the decision to likely fabricate in the United States, then ship it down. This decision then requires careful attention to the shipping & handling of the finished part, to ensure that the beveled edge does not sustain damage while in transit. The Camera team at SLAC has extensive experience shipping delicate hardware to Cerro Pachon (including the camera itself), making this a manageable hurdle.

### 4.5 Cost

When deciding whether or not to implement any change to an operational major observatory, the amount of impact that it has on the observation data is at the forefront of the decision. This 10 year survey is expected to produce over a petabyte of data per year, and will this change improve the quality of this immense amount of data? The Stray Light Working Group has found that it will. On a similar vein, for a project of this scale the manufacturing and labor costs of making and installing a simple, stationary baffle are insignificant, making them allowable expenses. Essentially, we can afford to have an L3 baffle made, and we have already proven to ourselves that it would be both beneficial and possible to add to the LSSTCam; however, the biggest question of cost is regarding the amount of downtime needed to have the baffle installed, because any amount of lost time on sky is a detriment to the survey. Considering this important distinction of the actual cost: time, the two main considerations are the difficulty of accessing the area and the time needed to install the baffle once access is gained.

Accessing this area of the LSSTCam while on the TMA is nontrivial. The region of the camera that would eventually house this baffle is very tightly packed with hardware, some of which would need to be removed to gain access, and other items which are impossible to remove in-situ. The Auto Changer (a large component of the Filter Exchange System) would need to be removed and stored on its cart for the duration of this work, then the Shutter would also need the same treatment. While the Auto Changer removal requires a crane lift and some fixtures, the Shutter removal requires use of a large, motorized tool called the Shutter Extraction Rail (pictured holding the shutter at right in Figure 20). The process to install and use the Shutter Extraction Rail is nontrivial and takes a significant amount of time. Despite both components being designed to be removable for maintenance reasons, the time needed to remove and reinstall these two major components would likely require three or more weeks off sky.

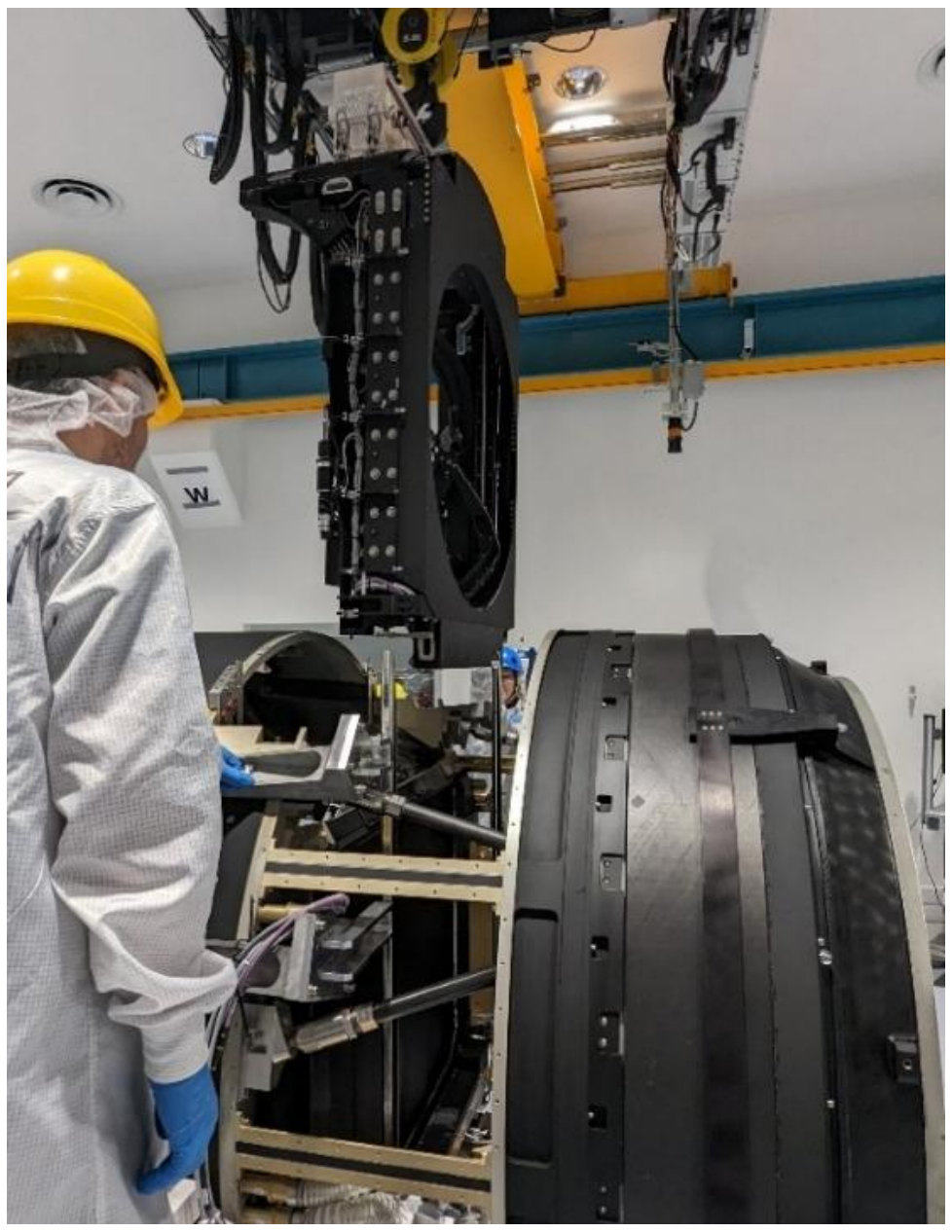

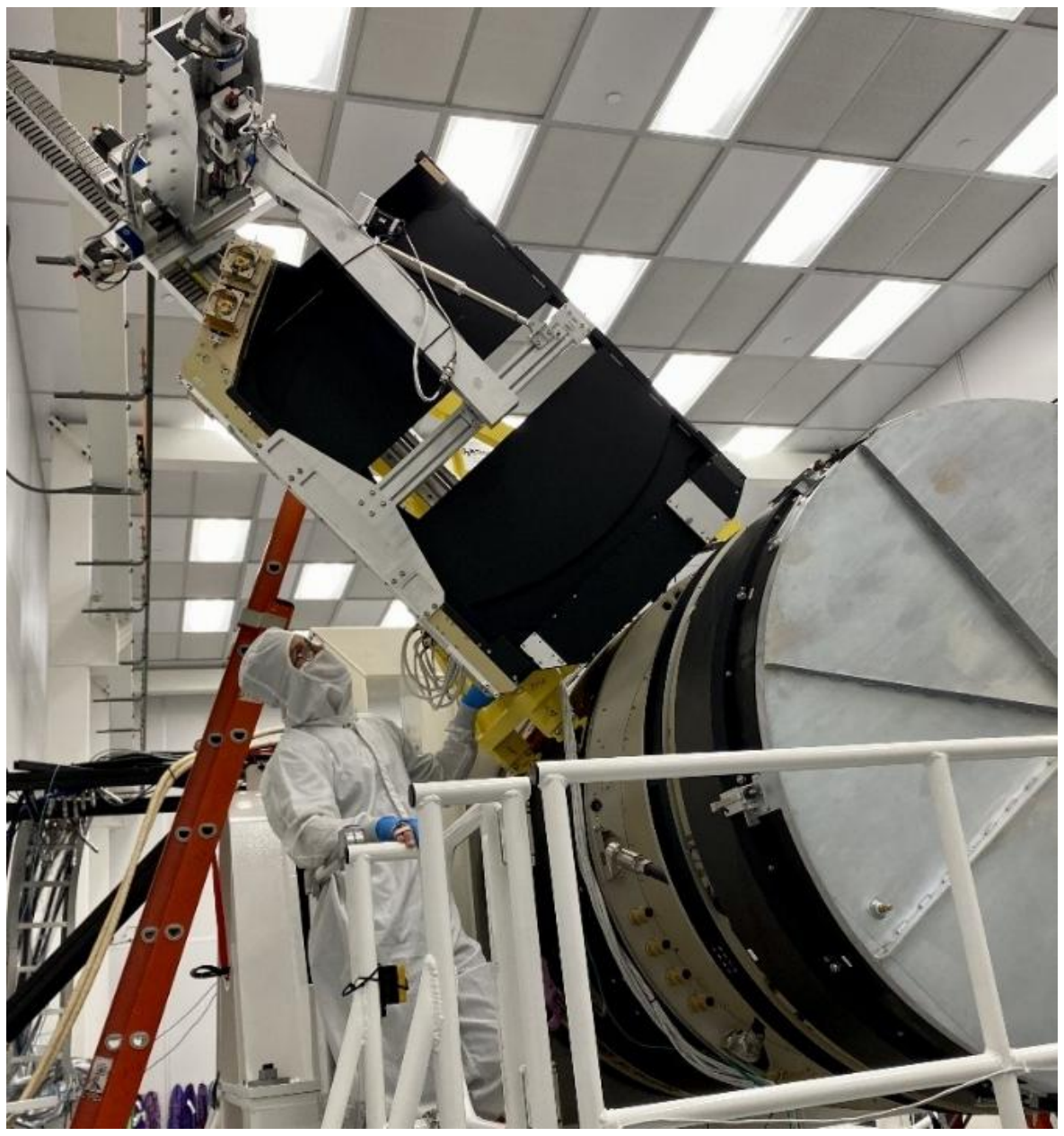

Figure 21: At the left is a photograph of the Auto Changer being installed into the LSSTCam in the clean room at the Vera Rubin Observatory, and at the right is a photograph of the shutter being installed in the cleanroom at SLAC. These images give an idea of the scale of operations needed to remove and install each of these pieces of hardware, and there is the added difficulty of these activities now having to take place on the TMA with workers in fall protection on the personnel platforms.

On the other hand, the L1-L2 lens assembly is impossible to remove while the camera is mounted on the telescope, meaning that the team must work around its 6 delicate carbon fiber mounting struts to thread the L3 baffle into its desired mounting location. Also, the L3 lens will not be covered up, so the exposed glass of the lens must not be contacted. This is doable but will need to be done with great care to prevent damage to the lens assembly, see Figure 21 below for an idea of the access on the TMA and one possible area of approach for installing this baffle into the LSSTCam.

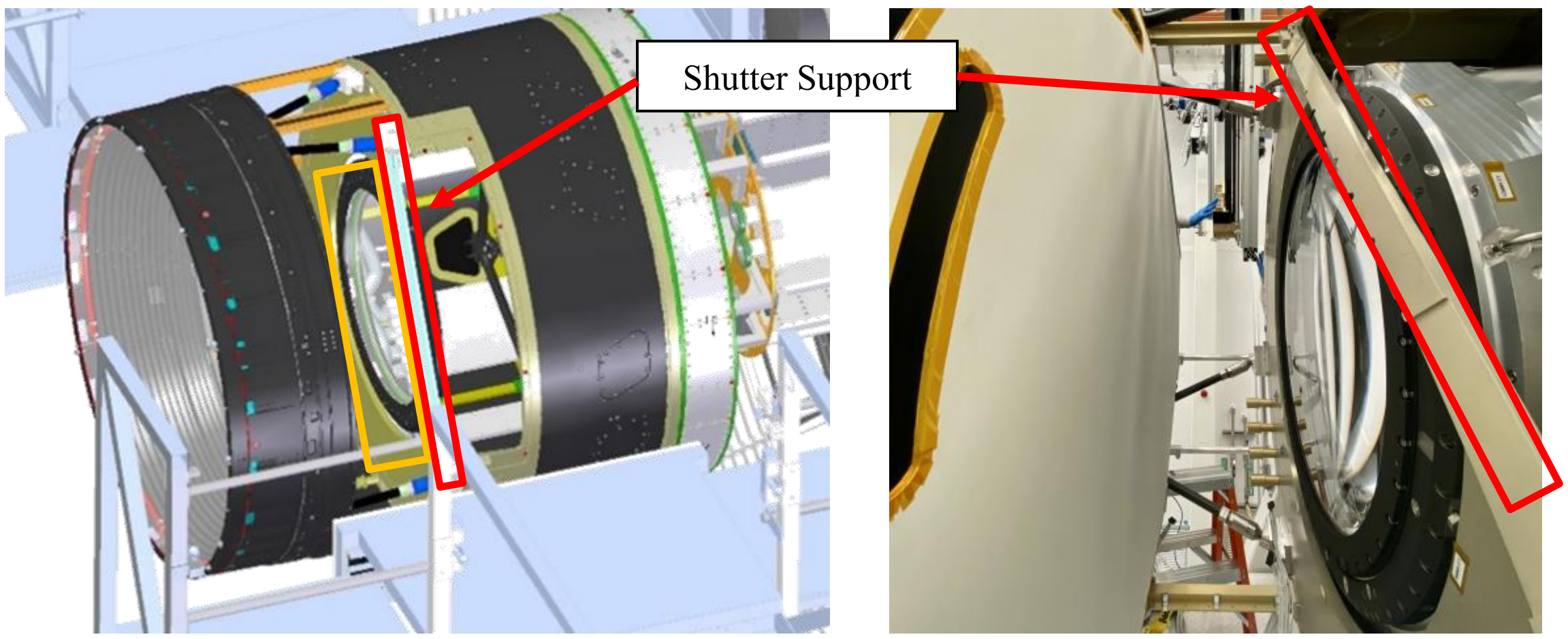

Figure 22: Close-up of Figure 8 at left with orange box showing potential entrance for installation. Photo from SLAC cleanroom at right showing even closer view of the real hardware in the appropriate configuration (Auto Changer and Shutter removed, L3 exposed), with shutter support bar labeled and boxed in red to show connection between two views.

The installation of the baffle is simple in nature and is anticipated to require much less time than the removal and re-installation of aforementioned larger mechanical assemblies, at a day or less of work time. However, the team would not recommend taking three weeks off sky just to install this baffle, the proposal would instead be to do it at a time when the observatory is already planning a swap of the shutter units for maintenance (planned to take place in the next year or two). This would consolidate downtime, with the only added time "cost" of baffle installation being a maximum of one night off sky, lumped in with an existing planned maintenance period.

## 5. CONCLUSIONS

Stray light is a bothersome and persistent problem especially in wide-field astronomy, but the systematic identification and mitigation work of the Rubin stray light working group throughout commissioning and into operations has begun to actively address each artifact. The ray tracing and mechanical analysis outlined in this paper have allowed the team to develop a baffle design that mitigates both the L3 Angel Wings and Horseshoe artifacts originating from the L3 chamfer.

The envisioned "Mickey Mouse" L3 Baffle design consists of corner cutouts that were shaped directly by the ray tracing results to prevent any additional vignetting while still blocking as many stray light paths as possible. The baffle edge bevel and optical black finish work together to minimize low order reflections from the baffle itself, and the mounting location allows for the execution of this baffle in the camera body.

While there are some final analyses to be completed and thus details to be solidified with respect to the baffle, the stray light working group is actively working on proving the potential benefit that adding a baffle to the LSSTCam at the L3 interface could provide. Future work includes the completion of the purge air CFD analysis, followed by any design tweaks and final Zemax analysis to confirm the performance of the baffle. Once this work is completed, the project shall decide whether to move forward with the fabrication and installation of the L3 baffle.

## ACKNOWLEDGMENTS

The authors of this paper would like to acknowledge the work of John Ku, whose contributions to the LSST Camera, especially the shutter system and surrounding region, served as a foundation for this paper, specifically supporting the ongoing fluid analysis. We dedicate this effort to his memory.

This material is based upon work supported in part by the National Science Foundation through Cooperative Agreements AST-1258333 and AST-2241526 and Cooperative Support Agreements AST-1202910 and 2211468 managed by the Association of Universities for Research in Astronomy (AURA), and the Department of Energy under Contract No. DE-AC02-76SF00515 with the SLAC National Accelerator Laboratory managed by Stanford University. Additional Rubin Observatory funding comes from private donations, grants to universities, and in-kind support from LSST-DA Institutional Members.

The authors also acknowledge that the Stanford AI Playground, specifically Claude (Anthropic) and Gemini (Google), was utilized for assistance in formatting and reviewing the language used in this paper (spelling, grammar, style). All engineering work and scientific results, as well as the analysis and conclusions from these results, are solely the work of the authors.

## REFERENCES

[1] Rodeghiero, G. et al., “An overview of the stray light findings and interpretation during the LSSTCam commissioning on-sky,” Proc. SPIE 14147, Ground-based and Airborne Telescopes XI, 14147-146 (7 July 2026)

[2] Rodeghiero, G. et al., (2025, July 28 - August 1). “Rubin Stray Light Working Group”, Rubin Community Workshop 2025, Tucson, AZ., United States.

[3] Flaugher, B. et al., "The Dark Energy Camera," The Astronomical Journal, Volume 150, Number 5, (2015). https://doi.org/10.1088/0004-6256/150/5/150

[4] Rodeghiero, G. et al., "Rubin stray light (& ghost) working group" [unpublished presentation], presented at Stray Light Working Group Meeting, Zoom, December 1, 2025.

[5] Drlica-Wagner, A. et al., “Study of a large-angle off-axis stray light path in Rubin Observatory commissioning,” Proc. SPIE 14147, Ground-based and Airborne Telescopes XI, 14147-121 (6 July 2026)

[6] Ku, J., Nordby, M., “L3 Region Heat Transfer Combined Analysis Report”, LCA-20252, SLAC, https://ls.st/lca-20252 (2023)

[7] Lange, T et al., "LSST camera raft integration support equipment: design, assembly, and test status," Proc. SPIE 10702, Ground-based and Airborne Instrumentation for Astronomy VII, 107024J (6 July 2018); https://doi.org/10.1117/12.2310061

[8] Rodeghiero, G. et al., "Rubin stray light (& ghost) working group" [unpublished presentation], presented at Stray Light Working Group Meeting, Zoom, February 23, 2026.

[9] Rodeghiero, G. et al., "Rubin stray light (& ghost) working group" [unpublished presentation], presented at Stray Light Working Group Meeting, Zoom, April 6, 2026.

[10] E. C. Fest, "Chapter 9 - Baffle and Cold Shield Design," in Stray Light Analysis and Control, SPIE Press, Bellingham, WA, pp. 163-182 (2013).

[11] “Selecting the Right Chamfer Cutter Tip Geometry,” Harvey Performance Company, https://www.harveyperformance.com/in-the-loupe/selecting-the-right-chamfer-cutter-tip-geometry (accessed 7 April 2026).

[12] “Wire EDM Machining Explained: How It Works and When to Use It,” Innovent Technologies, https://innoventtech.com/blog/wire-edm-machining-explained-how-it-works-and-when-to-use-it (accessed 7 April 2026).

[13] “Musou Black Color Product Information Page,” MusouBlack.com, https://www.musoublack.com/collections/acrylcolor/products/musou-black?variant=44798745346294 (accessed 23 March 2026)

[14] “Optical Black Product Information Page,” Pioneer Metal Finishing, https://www.pioneermetal.com/our-processes/proprietary-finishing-services/#optical-black (accessed 23 March 2026)

[15] “Vacuum Black Coating Product Information Page,” Acktar Advanced Coatings, https://acktar.com/product/vacuum-black/ (accessed 23 March 2026)

[16] “Substrates - Aeroglaze/Chemglaze Application Guide,” Socomore, https://store-us.socomore.com/substrates/ (accessed 23 March 2026)